\documentclass[prd,aps,showpacs,epsf,floats,10pt]{revtex4}
\usepackage{amssymb}
\usepackage{amsfonts}
\usepackage{amsmath}

\input{tcilatex}
\begin{document}

\title{Quantifying The Complexity Of Geodesic Paths On Curved Statistical
Manifolds Through Information Geometric Entropies and Jacobi Fields }
\author{Carlo Cafaro}
\email{carlo.cafaro@unicam.it}
\affiliation{Dipartimento di Fisica, Universit\`{a} di Camerino, I-62032 Camerino, Italy}
\author{Stefano Mancini}
\email{stefano.mancini@unicam.it}
\affiliation{Dipartimento di Fisica, Universit\`{a} di Camerino, I-62032 Camerino, Italy}

\begin{abstract}
We characterize the complexity of geodesic paths on a curved statistical
manifold $\mathcal{M}_{s}$ through the asymptotic computation of the
information geometric complexity $\mathcal{V}_{\mathcal{M}_{s}}$ and the
Jacobi vector field intensity $J_{\mathcal{M}_{s}}$. The manifold $\mathcal{M%
}_{s}$ is a $2l$-dimensional Gaussian model reproduced by an appropriate
embedding in a larger\textbf{\ }$4l$-dimensional Gaussian manifold and
endowed with a Fisher-Rao information metric $g_{\mu \nu }\left( \Theta
\right) $ with non-trivial off diagonal terms. These terms emerge due to the
presence of a correlational structure (embedding constraints) among the
statistical variables on the larger manifold and are characterized by
macroscopic correlational coefficients $r_{k}$.\textbf{\ }First, we observe
a power law decay of the information geometric complexity at a rate
determined by the coefficients $r_{k}$ and conclude that the non-trivial off
diagonal terms lead to the emergence of an asymptotic information geometric
compression of the explored macrostates $\Theta $ on $\mathcal{M}_{s}$.
Finally,\ we observe that the presence of such embedding constraints leads
to an attenuation of the asymptotic exponential divergence of the Jacobi
vector field intensity.
\end{abstract}

\maketitle

\section{\textbf{Introduction}}

Characterizing and understanding the mystery of the origin of life and the
unfolding of its evolution are perhaps the leading arguments motivating the
quantification of the extremely elusive concept of complexity \cite{L88,
GM95, F98}. Of course there are more pragmatic reasons that justify the
study of complexity, for example the problem of quantifying how complex is
quantum motion. This issue is of primary importance in quantum information
science. However, our knowledge of the relations between complexity,
dynamical stability, and chaoticity in a fully quantum domain is still not
satisfactory \cite{O98, C09}. The concept of complexity is very difficult to
define, its origin is not fully understood \cite{W84, W85, R95, H86, G86,
JPC89} and it is mainly for this reason that several quantitative measures
of complexity have appeared in the scientific literature \cite{L88, GM95,
F98}. In classical physics, measures of complexity are settled in a better
way. The Kolmogorov-Sinai metric entropy \cite{K65}, that is the sum of all
positive Lyapunov exponents \cite{P77}, is a powerful indicator of
unpredictability in classical systems and it measures the algorithmic
complexity of classical trajectories \cite{B83, B00, S89, W78}. Other known
measures of complexity are the logical depth \cite{B90}, the thermodynamic
depth \cite{S88}, the computational complexity \cite{P94}, the stochastic
complexity \cite{R86} and many more. Ideally, a good definition of
complexity should be mathematically rigorous and intuitive at the same time
so that it allows to tackle complexity-related problems in computation
theory and statistical physics as well. Of course, a quantitative measure of
complexity is truly useful if its range of applicability is not limited to
few unrealistic applications. It is also for this reason that in order to
properly define complexity measures, it should be clearly stated the reasons
why one is defining such a measure and what it is intended to capture.

It is known that classical complex systems exhibit local exponential
instability and are characterized by positive Lyapunov exponents \cite%
{alekseev}. Furthermore, the study of the relationship between entropy and
the complexity of trajectories of a dynamical system has always been an
active field of research \cite{B83, B00, S89}. In particular, in \cite{B83}
it was shown that the algorithmic complexity of trajectories of points in a
dynamical system is asymptotically equal to the entropy of the system.

It is commonly accepted that one of the major goals of physics is modeling
and predicting natural phenomena using relevant information about the system
of interest. Taking this statement seriously, it is reasonable to expect
that the laws of physics should reflect the methods for manipulating
information. Indeed, the less controversial opposite point of view may be
considered where the laws of physics are used to manipulate information.
This is exactly the point of view adopted in quantum information science
where information is manipulated using the laws of quantum mechanics \cite%
{nielsen1}.

Here we make use of the so-called Entropic Dynamics (ED) \cite{caticha1} and
Information Geometrodynamical Approach to Chaos (IGAC) \cite{carlo-tesi,
carlo-CSF}. ED is a theoretical framework that arises from the combination
of inductive inference (Maximum Entropy Methods, \cite{caticha2, ariel,
caticha-giffin, adom}) and Information Geometry \cite{amari}. The most
intriguing question being pursued in ED stems from the possibility of
deriving dynamics from purely entropic arguments. This is clearly valuable
in circumstances where microscopic dynamics may be too far removed from the
phenomena of interest, such as in complex biological or ecological systems,
or where it may just be unknown or perhaps even nonexistent, as in
economics. The applicability of ED has been extended to temporally-complex
(chaotic) dynamical systems on curved statistical manifolds and relevant
measures of chaoticity of such an information geometrodynamical approach to
chaos have been identified \cite{carlo-tesi}. IGAC arises as a theoretical
framework to study chaos in informational geodesic flows describing
physical, biological or chemical \ systems. A geodesic on a curved
statistical manifold $\mathcal{M}_{S}$ represents the maximum probability
path a complex dynamical system explores in its evolution between initial
and final macrostates. Each point of the geodesic is parametrized by the
macroscopic dynamical variables\textbf{\ }$\left\{ \Theta \right\} $
defining the macrostate of the system. Furthermore, each macrostate is in a
one-to-one correspondence with the probability distribution $\left\{ p\left(
X|\Theta \right) \right\} $ representing the maximally probable description
of the system being considered. The quantity $X$ is a microstate of the
microspace $\mathcal{X}$. The set of macrostates forms the parameter space $%
\mathcal{D}_{\Theta }$ while the set of probability distributions forms the
statistical manifold $\mathcal{M}_{S}$. IGAC\ is the information geometric
analogue of conventional geometrodynamical approaches \cite{casetti, di bari}
where the classical configuration space $\Gamma _{E}$\ is being replaced by
a statistical manifold $\mathcal{M}_{S}$\ with the additional possibility of
considering chaotic dynamics arising from non conformally flat metrics (the
Jacobi metric is always conformally flat, instead). It is an information
geometric extension of the Jacobi geometrodynamics (the geometrization of a
Hamiltonian system by transforming it to a geodesic flow \cite{jacobi}). The
reformulation of dynamics in terms of a geodesic problem allows the
application of a wide range of well-known geometrical techniques in the
investigation of the solution space and properties of the equation of
motion. The power of the Jacobi reformulation is that all of the dynamical
information is collected into a single geometric object in which all the
available manifest symmetries are retained- the manifold on which geodesic
flow is induced. For example, integrability of the system is connected with
existence of Killing vectors and tensors on this manifold. The sensitive
dependence of trajectories on initial conditions, which is a key ingredient
of chaos, can be investigated from the equation of geodesic deviation. In
the Riemannian \cite{casetti} and Finslerian \cite{di bari} (a Finsler
metric is obtained from a Riemannian metric by relaxing the requirement that
the metric be quadratic on each tangent space) geometrodynamical approach to
chaos in classical Hamiltonian systems, an active field of research concerns
the possibility of finding a rigorous relation among the sectional
curvature, the Lyapunov exponents, and the Kolmogorov-Sinai dynamical
entropy \cite{kawabe}.

In this article, inspired by the work presented in \cite{cafaroPD}, using
statistical inference and information geometric techniques, we characterize%
\textbf{\ }the complexity of geodesic paths on a curved statistical manifold 
$M_{s}$ through the asymptotic computation of the information geometric
complexity $\mathcal{V}_{\mathcal{M}_{s}}$ and the Jacobi vector field $J_{%
\mathcal{M}_{s}}$. The manifold $\mathcal{M}_{s}$ is a $2l$-dimensional
Gaussian model reproduced by an appropriate embedding in a larger\textbf{\ }$%
4l$-dimensional Gaussian manifold and endowed with a Fisher-Rao information
metric $g_{\mu \nu }\left( \Theta \right) $ with non-trivial off diagonal
terms. These terms in the information metric on the embedded manifold emerge
due to the presence of a correlational structure (embedding constraints)
among the statistical variables on the larger manifold and are characterized
by macroscopic correlational coefficients $r_{k}$.\textbf{\ }First, we
observe a power law decay of the information geometric complexity\textbf{\ }$%
\mathcal{V}_{\mathcal{M}_{s}}\left( \tau \right) \approx \exp \left[ 
\mathcal{S}_{\mathcal{M}_{s}}\left( \tau \right) \right] $\textbf{\ }at a
rate determined by the coefficients $r_{k}$ and conclude that the
non-trivial off diagonal terms lead to the emergence of an asymptotic
information geometric compression of the explored macrostates $\Theta $ on $%
\mathcal{M}_{s}$. Finally,\ we observe that the presence of such embedding
constraints leads to an attenuation of the asymptotic exponential divergence
of the Jacobi vector field intensity.

The layout of this article is as follows. In Section II, we present few
remarks on the theoretical structure of the IGAC and outline few selected
applications concerning the complexity characterization of geodesic paths on
curved statistical manifolds. In Section III, we describe the $2l$%
-dimensional curved statistical model considered, the embedded Gaussian
model endowed with a Fisher-Rao information metric with non-trivial off
diagonal terms. In Section IV, we present the asymptotic computation of the
information geometric entropy $\mathcal{S}_{\mathcal{M}_{s}}\left( \tau
\right) $. We observe a power law decay of the information geometric
complexity $\mathcal{V}_{\mathcal{M}_{s}}\left( \tau \right) $ at a rate
determined by the coefficients\textbf{\ }$r_{k}$ and conclude that
non-trivial off diagonal terms lead to the emergence of an asymptotic
information geometric compression of the explored macrostates on the
statistical configuration manifold considered. In Section V, we present the
asymptotic computation of the Jacobi fields on $\mathcal{M}_{S}$. We observe
that the presence of the embedding constraints lead to an attenuation of the
asymptotic exponential divergence of the Jacobi field intensity. Finally, in
Section VI we present final remarks and suggest further research directions.

\section{On the IGAC: Remarks and Applications}

In this Section, we present few remarks on the theoretical structure of the
IGAC and outline few selected applications concerning the complexity
characterization of geodesic paths on curved statistical manifolds. A more
detailed review appears in \cite{carlo-tesi}.

\subsection{Remarks}

As stated in the Introduction, the IGAC arises as an information geometric
framework to study chaos and complexity in informational geodesic flows
describing physical systems. A geodesic on a curved statistical manifold
represents the maximum probability path a complex dynamical system explores
in its evolution between initial and final macrostates $\Theta _{i}$ and $%
\Theta _{f}$, respectively. Each point of the geodesic on a $4l$-dimensional
statistical manifold represents a macrostate $\Theta $ parametrized by the
macroscopic dynamical variables $\Theta \equiv \left( \theta _{1}\text{,..., 
}\theta _{4l}\right) $ defining the macrostate of the system. Furthermore,
each macrostate is in a one-to-one correspondence with the probability
distribution $P\left( X|\Theta \right) $ representing the maximally probable
description of the system being considered. The set of macrostates forms the
parameter space while the set of probability distributions form the
statistical manifold. In what follows, we schematically outline the main
features underlying the construction of an arbitrary form of entropic
dynamics. First, the microstates of the system under investigation must be
defined. For the sake of reasoning, we assume the system is characterized by
an\textbf{\ }$2l$-dimensional microspace with microstates $X\equiv \left(
x_{1}\text{,..., }x_{2l}\right) $. The main goal of an ED model is that of
inferring "\emph{macroscopic predictions}" in the absence of detailed
knowledge of the microscopic nature of the arbitrary complex systems being
considered. More explicitly, by "macroscopic prediction" we mean knowledge
of the statistical parameters (expectation values) of the probability
distribution function that best reflects what is known about the system.
This is an important conceptual point. The probability distribution reflects
the system in general, not the microstates. Once the microstates have been
defined, we then select the relevant information about the system. In other
words, we have to select the macrospace of the system.

In general, one is given a manifold of probability distributions arising
from the maximum entropy formalism where distributions arise from the
maximization of the logarithmic relative entropy subjected to some
statements concerning averages (\emph{information constraints}). Given the
manifold of probability distributions, the (\emph{direct}) problem is to
find the corresponding Fisher-Rao information metric. However, not all
probability distributions are generated in this way. For instance,
probability distributions may emerge as a result of a change of variable
technique (parametric transformation law) \cite{tribus}. Furthermore, in
order to do physics, we are also concerned with the following (\emph{inverse}%
) problem: we want to design statistical manifolds with appropriate
geometries \cite{caticha-cafaro}.

\subsection{Applications}

In the following, we outline few selected applications concerning the
complexity characterization of geodesic paths on curved statistical
manifolds.

\subsubsection{Gaussian Statistical Models in the Absence of Correlations}

In \cite{cafaroPD}, we apply the IGAC to study the dynamics of a system with 
$l$ degrees of freedom, each one described by two pieces of relevant
information, its mean expected value and its variance (Gaussian statistical
macrostates). This leads to consider a statistical model on a non-maximally
symmetric $2l$-dimensional statistical manifold $\mathcal{M}_{s}$. It is
shown that $\mathcal{M}_{s}$ possesses a constant negative scalar curvature
proportional to the number of degrees of freedom of the system, $\mathcal{R}%
_{\mathcal{M}_{s}}=-l$. It is found that the system explores statistical
volume elements on $\mathcal{M}_{s}$ at an exponential rate. The information
geometric entropy $\mathcal{S}_{\mathcal{M}_{s}}$\ increases linearly in
time (statistical evolution parameter) and, moreover, is proportional to the
number of degrees of freedom of the system, $\mathcal{S}_{\mathcal{M}_{s}}$ $%
\overset{\tau \rightarrow \infty }{\sim }l\lambda \tau $ where $\lambda $ is
the maximum positive Lyapunov exponent characterizing the model. The
asymptotic linear information geometric entropy growth may be considered an
information-geometric analogue of the von Neumann entropy growth introduced
by Zurek-Paz, a \textit{quantum} feature of chaos. The geodesics on $%
\mathcal{M}_{s}$ are hyperbolic trajectories. Using the Jacobi-Levi-Civita
(JLC) equation for geodesic spread, we show that the Jacobi vector field
intensity $J_{\mathcal{M}_{s}}$ diverges exponentially and is proportional
to the number of degrees of freedom of the system, $J_{\mathcal{M}_{s}}$ $%
\overset{\tau \rightarrow \infty }{\sim }l\exp \left( \lambda \tau \right) $%
. The exponential divergence of the Jacobi vector field intensity $J_{%
\mathcal{M}_{s}}$ is a \textit{classical} feature of chaos. Therefore, we
conclude \ that $\mathcal{R}_{\mathcal{M}_{s}}=-l$, $J_{\mathcal{M}_{s}}%
\overset{\tau \rightarrow \infty }{\sim }l\exp \left( \lambda \tau \right) $
and $\mathcal{S}_{\mathcal{M}_{s}}\overset{\tau \rightarrow \infty }{\sim }%
l\lambda \tau $. Thus, $\mathcal{R}_{\mathcal{M}_{s}}$, $\mathcal{S}_{%
\mathcal{M}_{s}}$ and $J_{\mathcal{M}_{s}}$ behave as proper indicators of
chaoticity and are proportional to the number of Gaussian-distributed
microstates of the system. This proportionality, even though proven in a
very special case, leads to conclude there may be a substantial link among
these information geometric indicators of chaoticity.

\subsubsection{Gaussian Statistical Models in the Presence of Correlations
between Microvariables}

In \cite{carloPA2010}, we apply the IGAC to study the information
constrained dynamics of a system with $l=2$ microscopic degrees of freedom.
As working hypothesis, we assume that such degrees of freedom are
represented by two correlated Gaussian-distributed microvariables
characterized by the same variance. We show that the presence of
microcorrelations lead to the emergence of an asymptotic information
geometric compression of the statistical macrostates explored by the system
at a faster rate than that observed in absence of microcorrelations. This
result constitutes an important and explicit connection between
(micro)-correlations and (macro)-complexity in statistical dynamical
systems. The relevance of our finding is twofold: first, it provides a neat
description of the effect of information encoded in microscopic variables on
experimentally observable quantities defined in terms of dynamical
macroscopic variables; second, it clearly shows the change in behavior of
the macroscopic complexity of a statistical model caused by the existence of
correlations at the underlying microscopic level.

\subsubsection{Ensemble of Random Frequency Macroscopic Inverted Harmonic
Oscillators}

\emph{\ }In \cite{caticha-cafaro}, we explore the possibility of using well
established principles of inference to derive Newtonian dynamics from
relevant prior information codified into an appropriate statistical
manifold. The basic assumption is that there is an irreducible uncertainty
in the location of particles so that the state of a particle is defined by a
probability distribution. The corresponding configuration space is a
statistical manifold the geometry of which is defined by the Fisher-Rao
information metric. The trajectory follows from a principle of inference,
the method of Maximum Entropy. There is no need for additional "physical"
postulates such as an action principle or equation of motion, nor for the
concept of mass, momentum and of phase space, not even the notion of time.
The resulting "entropic" dynamics reproduces Newton's mechanics for any
number of particles interacting among themselves and with external fields.
Both the mass of the particles and their interactions are explained as a
consequence of the underlying statistical manifold.

Following this line of reasoning, in \cite{carlo-CSF, EJTP} we present an
information geometric analogue of the Zurek-Paz quantum chaos criterion in
the \textit{classical reversible limit}. This analogy is illustrated by
applying the IGAC to a set of\textbf{\ }$n$\textbf{-}uncoupled
three-dimensional anisotropic inverted harmonic oscillators characterized by
a Ohmic distributed frequency spectrum.

\subsubsection{IGAC of regular and chaotic quantum spin chains}

In \cite{cafaroMPLB, cafaroPA}, we study the entropic dynamics on curved
statistical manifolds induced by classical probability distributions of
common use in the study of regular and chaotic quantum energy level
statistics. Specifically, we propose an information geometric
characterization of chaotic (integrable) energy level statistics of a
quantum antiferromagnetic Ising spin chain in a tilted (transverse) external
magnetic field. We consider the IGAC of a Poisson distribution coupled to an
Exponential bath (spin chain in a \textit{transverse} magnetic field,
regular case) and that of a Wigner-Dyson distribution coupled to a Gaussian
bath (spin chain in a \textit{tilted} magnetic field, chaotic case).
Remarkably, we show that in the former case the IGE exhibits asymptotic
logarithmic growth while in the latter case the IGE exhibits asymptotic
linear growth.

\section{The Model}

In this Section, we emphasize the main reasoning underlying the choice of
the new proposed statistical model and study its information geometric
properties.

\subsection{Motivations}

We want to make reliable macroscopic predictions when only partial knowledge
on the micro-structure of a system is available. As stated in the
Introduction, the complexity of such predictions is quantified in terms of
the IGE and the Jacobi field intensity. In this manuscript we seek an answer
to the following question: isn't simpler to make macroscopic predictions
when the available pieces of information are not independent? Stated
otherwise, does an increase in the correlational structure of the dynamical
equations for the statistical variables labelling a macrostate of a system
imply a reduction in the complexity of the geodesic paths? It is reasonable
to expect that the emergence of a correlational structure in the form of
constraints among the variables labelling the macrostates of a system would
lead to a highly constrained dynamics and, consequently, to a reduction in
the complexity of making macroscopic predictions. In what follows, we
attempt to give a quantitative answer to the above-mentioned questions.

We propose to compare the complexity of making predictions in two different
scenarios. In the first scenario ($4l$-dimensional larger Gaussian model) ,
we assume a system with $2l$ degrees of freedom $x_{k}$, each one being
Gaussian distributed. The probability distribution describing the whole
system is given by,%
\begin{equation}
P\left( X|\Theta \right) =\dprod\limits_{k=1}^{2l}p\left( x_{k}|\mu _{k}%
\text{, }\sigma _{k}\right) \text{,}  \label{scenario1}
\end{equation}%
where $X\equiv \left( x_{1}\text{,..., }x_{2l}\right) $, $\Theta \equiv
\left( \mu _{1}\text{,..., }\mu _{2l}\text{, }\sigma _{1}\text{,..., }\sigma
_{2l}\right) $ and $p\left( x_{k}|\mu _{k}\text{, }\sigma _{k}\right) $ is
defined as,%
\begin{equation}
p\left( x_{k}|\mu _{k}\text{, }\sigma _{k}\right) \overset{\text{def}}{=}%
\frac{1}{\sqrt{2\pi \sigma _{k}^{2}}}\exp \left[ -\frac{\left( x_{k}-\mu
_{k}\right) ^{2}}{2\sigma _{k}^{2}}\right] \text{.}
\end{equation}%
In \cite{cafaroPD, cafaroIJTP}, we provided an analytical estimate for the
complexity of geodesic paths on the statistical manifolds of distributions
given in (\ref{scenario1}). In the second scenario ($2l$-dimensional
embedded Gaussian model), we consider the very same system with $2l$ degrees
of freedom $x_{k}$. However, the microvariables $x_{k}$ are described by
probability distributions characterized by statistical variables subject to
a set of \ $2l$-constraints,%
\begin{equation}
\sigma _{2j}=\sigma _{2j-1}\text{ and, }\mu _{2j}=\mu _{2j}\left( \mu _{2j-1}%
\text{, }\sigma _{2j-1}\right) \text{ (embedding constraints),}
\label{correlational}
\end{equation}%
with $j=1$,..., $l$. Therefore, the probability distribution describing the
whole system in this second scenario becomes,%
\begin{equation}
\tilde{P}\left( X|\tilde{\Theta}\right) =\dprod\limits_{j=1}^{l}\tilde{p}%
\left( x_{2j-1}\text{, }x_{2j}|\mu _{2j-1}\text{, }\sigma _{2j-1}\right) 
\text{,}  \label{scenario2}
\end{equation}%
where $X\equiv \left( x_{1}\text{,..., }x_{2l}\right) $, $\tilde{\Theta}%
\equiv \left( \mu _{1}\text{, }\mu _{3}\text{,..., }\mu _{2l-1}\text{, }%
\sigma _{1}\text{, }\sigma _{3}\text{,..., }\sigma _{2l-1}\right) $ and $%
\tilde{p}\left( x_{2j-1}\text{, }x_{2j}|\mu _{2j-1}\text{, }\sigma
_{2j-1}\right) $ is defined as,%
\begin{equation}
\tilde{p}\left( x_{2j-1}\text{, }x_{2j}|\mu _{2j-1}\text{, }\sigma
_{2j-1}\right) \overset{\text{def}}{=}\frac{1}{2\pi \sigma _{2j-1}^{2}}\exp %
\left[ -\frac{\left( x_{2j-1}-\mu _{2j-1}\right) ^{2}+\left[ x_{2j}-\mu
_{2j}\left( \mu _{2j-1}\text{, }\sigma _{2j-1}\right) \right] ^{2}}{2\sigma
_{2j-1}^{2}}\right] \text{,}
\end{equation}%
with $j=1$,..., $l$. Our purpose is computing the complexity of geodesic
paths on the $2l$-dimensional statistical manifold $\mathcal{M}_{S}^{\left( 
\text{embedded}\right) }\equiv \mathcal{M}_{S}=\left\{ \tilde{P}\left( X|%
\tilde{\Theta}\right) \right\} $ and compare it with the one obtained on the 
$4l$-dimensional manifold $\mathcal{M}_{S}^{\left( \text{larger}\right)
}=\left\{ P\left( X|\Theta \right) \right\} $. We expect that the emergence
of the correlational structure defined in (\ref{correlational}) between
pairs of macroscopic statistical variables will give rise to a reduction of
the system's complexity.

Except for an overall scale constant and a convenient re-scaling of
variables, the information metric on the $2l$-dimensional embedded manifold $%
\mathcal{M}_{S}$ is given by (see Appendix A for the explicit derivation),%
\begin{equation}
dS_{\mathcal{M}_{S}}^{2}=\dsum\limits_{j=1}^{l}\frac{1}{\sigma _{2j-1}^{2}}%
\left( d\mu _{2j-1}^{2}+2r_{2j-1}d\mu _{2j-1}d\sigma _{2j-1}+2d\sigma
_{2j-1}^{2}\right) \text{,}  \label{ntfm}
\end{equation}%
where the coefficients $r_{2j-1}$ are defined as,%
\begin{equation}
r_{2j-1}\overset{\text{def}}{=}\frac{\frac{\partial \mu _{2j}}{\partial \mu
_{2j-1}}\frac{\partial \mu _{2j}}{\partial \sigma _{2j-1}}}{\left[ 1+\left( 
\frac{\partial \mu _{2j}}{\partial \mu _{2j-1}}\right) ^{2}\right] ^{\frac{1%
}{2}}\left[ 2+\frac{1}{2}\left( \frac{\partial \mu _{2j}}{\partial \sigma
_{2j-1}}\right) ^{2}\right] ^{\frac{1}{2}}}  \label{rk}
\end{equation}%
The explicit expressions of such coefficients depend on the functional
parametric form given to the embedding constraints $\mu _{2j}=\mu
_{2j}\left( \mu _{2j-1}\text{, }\sigma _{2j-1}\right) $. More details are
given in Appendix A. From (\ref{rk}) it follows that the coefficients $%
r_{2j-1}$ are non-zero if and only if $\mu _{2j}$ depends on both $\mu
_{2j-1}$ and $\sigma _{2j-1}$. Therefore, we may conclude that the emergence
of the non-trivial off-diagonal terms in (\ref{ntfm}) is a consequence of
the previously mentioned correlational structure arising from the embedding
constraints. Motivated by these considerations, we will name from now on the
coefficients $r_{2j-1}$, \emph{macroscopic correlational coefficients}.

\subsection{Information Geometry of the Model}

In this Subsection, we discuss the main steps leading to computation of the
asymptotic temporal behavior of the dynamical complexity of geodesic
trajectories for the $2l$-dimensional Gaussian statistical model. For the
sake of notational simplicity and in view of the involved analysis that we
will present shortly, we replace $\tilde{\Theta}$ in (\ref{scenario2}) with $%
\Theta \equiv \left( \mu _{1}\text{,..., }\mu _{l}\text{, }\sigma _{1}\text{%
,..., }\sigma _{l}\right) $ and $\tilde{P}\left( X|\tilde{\Theta}\right) $
in (\ref{scenario2}) with $P\left( X|\Theta \right) $ so that $\mathcal{M}%
_{S}^{\left( \text{embedded}\right) }\equiv \mathcal{M}_{S}=\left\{ P\left(
X|\Theta \right) \right\} $. We begin to study the IGAC arising from the
Fisher-Rao metric defined as,%
\begin{equation}
ds_{\mathcal{M}_{s}}^{2}\overset{\text{def}}{=}\text{ }g_{ij}\left( \Theta
\right) d\Theta ^{i}d\Theta ^{j}=\sum_{k=1}^{l}\left( \frac{1}{\sigma
_{k}^{2}}d\mu _{k}^{2}+\frac{2r_{k}}{\sigma _{k}^{2}}d\mu _{k}d\sigma _{k}+%
\frac{2}{\sigma _{k}^{2}}d\sigma _{k}^{2}\right) \text{, with }i\text{, }j=1%
\text{,..., }2l\text{.}  \label{SM}
\end{equation}%
We assume positive macroscopic correlational coefficients $r_{k}\in \left( 0%
\text{, }1\right) $, $\forall k=1$,..., $l$. The Fisher-Rao metric tensor $%
g_{ij}\left( \Theta \right) \overset{\text{def}}{=}g_{ij}\left( \mu _{1}%
\text{,...}\mu _{l}\text{; }\sigma _{1}\text{,..., }\sigma _{l}\right) $
leading to the line element in (\ref{SM}) is given by,%
\begin{equation}
\left[ g_{ij}\left( \Theta \right) \right] _{2l\times 2l}=\left( 
\begin{array}{cccc}
M_{2\times 2}^{\left( 1\right) } & 0 & 0 & 0 \\ 
0 & \cdot & 0 & 0 \\ 
0 & 0 & \cdot & 0 \\ 
0 & 0 & 0 & M_{2\times 2}^{\left( l\right) }%
\end{array}%
\right) \text{, with }i\text{, }j=1\text{,..., }2l\text{.}  \label{here}
\end{equation}%
where $M_{2\times 2}^{\left( k\right) }$ is the two-dimensional matrix
defined as,%
\begin{equation}
\left[ M_{2\times 2}^{\left( k\right) }\right] \overset{\text{def}}{=}\frac{1%
}{\sigma _{k}^{2}}\left( 
\begin{array}{cc}
1 & r_{k} \\ 
r_{k} & 2%
\end{array}%
\right) \text{ with }k=1\text{,..., }l\text{.}
\end{equation}%
The inverse matrix $\left[ M_{2\times 2}^{\left( k\right) }\right] ^{-1}$,
useful for computing the Christoffel connection coefficients and other
quantities characterizing the information geometry of $\mathcal{M}_{s}$ is
given by,%
\begin{equation}
\left[ M_{2\times 2}^{\left( k\right) }\right] ^{-1}\overset{\text{def}}{=}%
\frac{\sigma _{k}^{2}}{2-r_{k}^{2}}\left( 
\begin{array}{cc}
2 & -r_{k} \\ 
-r_{k} & 1%
\end{array}%
\right) \text{ with }k=1\text{,..., }l\text{.}
\end{equation}%
It can be shown \cite{cafaro-mancini-1} that the scalar curvature of such $%
2l $-dimensional manifold is given by,%
\begin{equation}
\mathcal{R}_{\mathcal{M}_{s}}\left( r_{1}\text{,..., }r_{l}\right)
=-2\sum_{k=1}^{l}\left( 2-r_{k}^{2}\right) ^{-1}\text{.}
\end{equation}%
Notice that in the limit of vanishing coefficients\textbf{\ }$\left\{
r_{k}=0\right\} $, $\mathcal{R}_{\mathcal{M}_{s}}=-l$ as shown in \cite%
{cafaroIJTP}. The computation of geodesic equations on the $2l$-dimensional
Gaussian statistical manifold $\mathcal{M}_{s}$ leads to the following
coupled systems of nonlinear second order ordinary differential equations,%
\begin{eqnarray}
0 &=&\frac{d^{2}\mu _{k}}{d\tau ^{2}}-\frac{r_{k}}{2-r_{k}^{2}}\frac{1}{%
\sigma _{k}}\left( \frac{d\mu _{k}}{d\tau }\right) ^{2}-\frac{4}{2-r_{k}^{2}}%
\frac{1}{\sigma _{k}}\frac{d\mu _{k}}{d\tau }\frac{d\sigma _{k}}{d\tau }-%
\frac{2r_{k}}{2-r_{k}^{2}}\frac{1}{\sigma _{k}}\left( \frac{d\sigma _{k}}{%
d\tau }\right) ^{2}\text{,}  \notag \\
&&  \notag \\
0 &=&\frac{d^{2}\sigma _{k}}{d\tau ^{2}}+\frac{1}{2-r_{k}^{2}}\frac{1}{%
\sigma _{k}}\left( \frac{d\mu _{k}}{d\tau }\right) ^{2}+\frac{2r_{k}}{%
2-r_{k}^{2}}\frac{1}{\sigma _{k}}\frac{d\mu _{k}}{d\tau }\frac{d\sigma _{k}}{%
d\tau }+\frac{2r_{k}^{2}-2}{2-r_{k}^{2}}\frac{1}{\sigma _{k}}\left( \frac{%
d\sigma _{k}}{d\tau }\right) ^{2}\text{.}
\end{eqnarray}%
with $k=1$,..., $l$. When $r_{k}\rightarrow 0$, $\forall k$ we get the
ordinary Gaussian system of nonlinear and coupled ordinary differential
equations. Integration of such coupled system of nonlinear second order
ordinary differential equations in is highly non trivial. However, this
problem can be tackled using the information geometric diagonalization
procedure introduced in \cite{cafaro-mancini-1}. The information metric
tensor $\hat{g}\left( \mu _{1}\text{,...}\mu _{l}\text{; }\sigma _{1}\text{%
,..., }\sigma _{l}\right) \overset{\text{def}}{=}\hat{g}\left( \Theta
\right) $ in (\ref{SM}) is symmetric and therefore diagonalizable. The
eigenvalues of such matrix are,%
\begin{equation}
\alpha _{\pm }\left( r_{k}\right) \overset{\text{def}}{=}\frac{3\pm \sqrt{%
\Delta \left( r_{k}\right) }}{2}\text{, }\Delta \left( r_{k}\right)
=1+4r_{k}^{2}\text{, with }k=1\text{,..., }l\text{.}
\end{equation}%
The eigenvectors $\Theta _{+}^{\left( k\right) }\overset{\text{def}}{=}%
\Theta _{+}\left( r_{k}\right) $ and $\Theta _{-}^{\left( k\right) }\overset{%
\text{def}}{=}\Theta _{-}\left( r\right) $ corresponding to $\alpha
_{+}\left( r_{k}\right) $ and $\alpha _{-}\left( r_{k}\right) $,
respectively, are,%
\begin{equation}
\Theta _{+}\left( r_{k}\right) =\left( 
\begin{array}{c}
1 \\ 
\frac{1+\sqrt{\Delta \left( r_{k}\right) }}{2r_{k}}%
\end{array}%
\right) \text{and, }\Theta _{-}\left( r_{k}\right) =\left( 
\begin{array}{c}
1 \\ 
\frac{1-\sqrt{\Delta \left( r_{k}\right) }}{2r_{k}}%
\end{array}%
\right) \text{ with }k=1\text{,..., }l\text{. }
\end{equation}%
The diagonalized information matrix $\left[ \hat{g}^{\prime }\left( \Theta
\left( \tilde{\Theta}\right) \right) \right] _{\mathcal{B}_{\text{new}}}$ in
the new basis $\mathcal{B}_{\text{new}}$ satisfies the following relation, 
\begin{equation}
\left[ \hat{g}\left( \Theta \right) \right] _{\mathcal{B}_{\text{old}%
}}=E_{2l\times 2l}\left( r_{1}\text{,..., }r_{l}\right) \left[ \hat{g}%
^{\prime }\left( \Theta \left( \tilde{\Theta}\right) \right) \right] _{%
\mathcal{B}_{\text{new}}}E_{2l\times 2l}^{-1}\left( r_{1}\text{,..., }%
r_{l}\right) \text{,}
\end{equation}%
where, in an explicit way, we obtain%
\begin{equation}
\left[ \hat{g}^{\prime }\left( \Theta \left( \tilde{\Theta}\right) \right) %
\right] _{\mathcal{B}_{\text{new}}}=\left( 
\begin{array}{cccc}
D_{2\times 2}^{\left( 1\right) } & 0 & 0 & 0 \\ 
0 & \cdot & 0 & 0 \\ 
0 & 0 & \cdot & 0 \\ 
0 & 0 & 0 & D_{2\times 2}^{\left( l\right) }%
\end{array}%
\right) \text{,}
\end{equation}%
with the two-dimensional diagonal matrices $D_{2\times 2}^{\left( k\right) }$
defined as,%
\begin{equation}
\left[ D_{2\times 2}^{\left( k\right) }\right] \overset{\text{def}}{=}\frac{1%
}{\sigma _{k}^{2}\left( \tilde{\mu}_{k}\text{, }\tilde{\sigma}_{k}\right) }%
\left( 
\begin{array}{cc}
\frac{3-\sqrt{\Delta \left( r_{k}\right) }}{2} & 0 \\ 
0 & \frac{3+\sqrt{\Delta \left( r_{k}\right) }}{2}%
\end{array}%
\right) \text{ with }k=1\text{,..., }l\text{.}
\end{equation}%
The columns of the matrix $E_{2l\times 2l}\left( r_{1}\text{,..., }%
r_{l}\right) $ encode the eigenvectors of $\left[ \hat{g}\left( \Theta
\right) \right] _{\mathcal{B}_{\text{old}}}$, $\Theta _{+}^{\left( k\right)
} $ and $\Theta _{-}^{\left( k\right) }$ and is given by,%
\begin{equation}
\left[ E\left( r_{1}\text{,..., }r_{l}\right) \right] _{2l\times 2l}=\left( 
\begin{array}{cccc}
E_{2\times 2}^{\left( 1\right) } & 0 & 0 & 0 \\ 
0 & \cdot & 0 & 0 \\ 
0 & 0 & \cdot & 0 \\ 
0 & 0 & 0 & E_{2\times 2}^{\left( l\right) }%
\end{array}%
\right) \text{,}
\end{equation}%
where the two-dimensional matrices $E_{2\times 2}^{\left( k\right) }$ are,%
\begin{equation}
\left[ E_{2\times 2}^{\left( k\right) }\right] \overset{\text{def}}{=}\left( 
\begin{array}{cc}
1 & 1 \\ 
\frac{1-\sqrt{\Delta \left( r_{k}\right) }}{2r_{k}} & \frac{1+\sqrt{\Delta
\left( r_{k}\right) }}{2r_{k}}%
\end{array}%
\right) \text{.}  \label{ZZ}
\end{equation}%
The relevance of $E_{2l\times 2l}\left( r_{1}\text{,..., }r_{l}\right) $
(and its inverse) is in expressing the set of macrovariables $\left( \mu _{1}%
\text{,...}\mu _{l}\text{; }\sigma _{1}\text{,..., }\sigma _{l}\right) $ in
terms of the new statistical variables $\left( \tilde{\mu}_{1}\text{,..., }%
\tilde{\mu}_{l}\text{; }\tilde{\sigma}_{1}\text{,..., }\tilde{\sigma}%
_{l}\right) $,%
\begin{equation}
g_{ij}\left( \mu _{1}\text{,...}\mu _{l}\text{; }\sigma _{1}\text{,..., }%
\sigma _{l}\right) \overset{\text{diag}}{\longrightarrow }g_{ij}^{\prime
}\left( \tilde{\mu}_{1}\text{,..., }\tilde{\mu}_{l}\text{; }\tilde{\sigma}%
_{1}\text{,..., }\tilde{\sigma}_{l}\right) \text{.}
\end{equation}%
From differential geometry arguments \cite{LEE}, it follows that%
\begin{equation}
\left( 
\begin{array}{c}
\partial _{\tilde{\mu}_{1}} \\ 
\cdot \\ 
\cdot \\ 
\partial _{\tilde{\sigma}_{l}}%
\end{array}%
\right) =E_{2l\times 2l}\left( r_{1}\text{,..., }r_{l}\right) \left( 
\begin{array}{c}
\partial _{\mu _{1}} \\ 
\cdot \\ 
\cdot \\ 
\partial _{\sigma _{l}}%
\end{array}%
\right) \text{ and, }\left( 
\begin{array}{c}
\mu _{1} \\ 
\cdot \\ 
\cdot \\ 
\sigma _{l}%
\end{array}%
\right) =E_{2l\times 2l}\left( r_{1}\text{,..., }r_{k}\right) \left( 
\begin{array}{c}
\tilde{\mu}_{1} \\ 
\cdot \\ 
\cdot \\ 
\tilde{\sigma}_{l}%
\end{array}%
\right) \text{.}  \label{ZZZ}
\end{equation}%
Substituting (\ref{ZZ}) in (\ref{ZZZ}), we finally obtain the formal
relation between the old and new set of macrovariables labelling the $2l$%
-dimensional macrostates $\Theta $ of the embedded Gaussian statistical
model in presence of non-trivial off-diagonal terms,%
\begin{equation}
\mu _{k}\left( \tilde{\mu}_{k}\text{, }\tilde{\sigma}_{k}\right) \overset{%
\text{def}}{=}\tilde{\mu}_{k}+\tilde{\sigma}_{k}\text{ and, }\sigma \left( 
\tilde{\mu}_{k}\text{, }\tilde{\sigma}_{k}\right) \overset{\text{def}}{=}%
\frac{1-\sqrt{\Delta \left( r_{k}\right) }}{2r_{k}}\tilde{\mu}_{k}+\frac{1+%
\sqrt{\Delta \left( r_{k}\right) }}{2r_{k}}\text{ }\tilde{\sigma}_{k}\text{,
with }k=1\text{,..., }l\text{.}  \label{re}
\end{equation}%
After having introduced the information geometric diagonalization procedure,
the new line element $ds^{\prime 2}\left( \tilde{\mu}_{1}\text{,..., }\tilde{%
\mu}_{l}\text{; }\tilde{\sigma}_{1}\text{,..., }\tilde{\sigma}_{l}\right) $
to be considered becomes,%
\begin{equation}
ds^{\prime 2}\left( \tilde{\mu}_{1}\text{,..., }\tilde{\mu}_{l}\text{; }%
\tilde{\sigma}_{1}\text{,..., }\tilde{\sigma}_{l}\right) =\sum_{k=1}^{l}%
\left[ \frac{\alpha _{-}\left( r_{k}\right) }{\left[ a_{0}\left(
r_{k}\right) \tilde{\mu}_{k}+a_{1}\left( r_{k}\right) \tilde{\sigma}_{k}%
\right] ^{2}}d\tilde{\mu}_{k}^{2}+\frac{\alpha _{+}\left( r_{k}\right) }{%
\left[ a_{0}\left( r_{k}\right) \tilde{\mu}_{k}+a_{1}\left( r_{k}\right) 
\tilde{\sigma}_{k}\right] ^{2}}d\tilde{\sigma}_{k}^{2}\right] \text{,}
\end{equation}%
where,%
\begin{equation}
\alpha _{\pm }\left( r_{k}\right) \overset{\text{def}}{=}\frac{3\pm \sqrt{%
\Delta \left( r_{k}\right) }}{2}\text{, }a_{0}\left( r_{k}\right) \overset{%
\text{def}}{=}\frac{1-\sqrt{\Delta \left( r_{k}\right) }}{2r_{k}}\text{, }%
a_{1}\left( r_{k}\right) \overset{\text{def}}{=}\frac{1+\sqrt{\Delta \left(
r_{k}\right) }}{2r_{k}}\text{ and, }\Delta \left( r_{k}\right) \overset{%
\text{def}}{=}1+4r_{k}^{2}\text{.}  \label{joe}
\end{equation}%
Notice that $ds^{\prime 2}\left( \tilde{\mu}_{1}\text{,..., }\tilde{\mu}_{l}%
\text{; }\tilde{\sigma}_{1}\text{,..., }\tilde{\sigma}_{l}\right) $ can be
rewritten as,%
\begin{equation}
ds^{\prime 2}\left( \tilde{\mu}_{1}\text{,., }\tilde{\mu}_{l}\text{; }\tilde{%
\sigma}_{1}\text{,., }\tilde{\sigma}_{l}\right) =\sum_{k=1}^{l}\left[ \frac{%
\alpha _{-}\left( r_{k}\right) }{\left[ a_{1}\left( r_{k}\right) \right] ^{2}%
}\frac{1}{\tilde{\sigma}_{k}^{2}}\frac{1}{\left( 1+\frac{a_{0}\left(
r_{k}\right) }{a_{1}\left( r_{k}\right) }\frac{\tilde{\mu}_{k}}{\tilde{\sigma%
}_{k}}\right) ^{2}}d\tilde{\mu}_{k}^{2}+\frac{\alpha _{+}\left( r_{k}\right) 
}{\left[ a_{1}\left( r_{k}\right) \right] ^{2}}\frac{1}{\tilde{\sigma}%
_{k}^{2}}\frac{1}{\left( 1+\frac{a_{0}\left( r_{k}\right) }{a_{1}\left(
r_{k}\right) }\frac{\tilde{\mu}_{k}}{\tilde{\sigma}_{k}}\right) ^{2}}d\tilde{%
\sigma}_{k}^{2}\right] \text{.}
\end{equation}%
As a working hypothesis, we assume that $\frac{a_{0}\left( r_{k}\right) }{%
a_{1}\left( r_{k}\right) }\frac{\tilde{\mu}_{k}\left( \tau \right) }{\tilde{%
\sigma}_{k}\left( \tau \right) }\ll 1$ for $\tau \gg 1$ and for each $k=1$%
,.., $l$. Stated otherwise, we assume that%
\begin{equation}
\lim_{\tau \rightarrow \infty }\left[ \frac{\tilde{\mu}_{k}\left( \tau
\right) }{\tilde{\sigma}_{k}\left( \tau \right) }\right] \ll \min_{r\in
\left( 0\text{,}1\right) }\left\vert \frac{a_{1}\left( r_{k}\right) }{%
a_{0}\left( r_{k}\right) }\right\vert =\min_{r_{k}\in \left( 0\text{,}%
1\right) }\left\vert \frac{1+\sqrt{1+4r_{k}^{2}}}{1-\sqrt{1+4r_{k}^{2}}}%
\right\vert \simeq 2.6\text{.}  \label{1}
\end{equation}%
Then, in the\textit{\ asymptotic long-time limit }\cite{boris}, the notion
of distinguishability between probability distributions on the diagonalized
statistical manifold is quantified by the following line element,%
\begin{equation}
ds^{\prime 2}\left( \tilde{\mu}_{1}\text{,., }\tilde{\mu}_{l}\text{; }\tilde{%
\sigma}_{1}\text{,., }\tilde{\sigma}_{l}\right) =\frac{\alpha _{-}\left(
r_{k}\right) }{\left[ a_{1}\left( r_{k}\right) \right] ^{2}}\frac{1}{\tilde{%
\sigma}_{k}^{2}}d\tilde{\mu}_{k}^{2}+\frac{\alpha _{+}\left( r_{k}\right) }{%
\left[ a_{1}\left( r_{k}\right) \right] ^{2}}\frac{1}{\tilde{\sigma}_{k}^{2}}%
d\tilde{\sigma}_{k}^{2}\text{.}  \label{qq}
\end{equation}%
Recall that the Christoffel connection coefficients $\Gamma _{ij}^{n}$ are
defined as,%
\begin{equation}
\Gamma _{ij}^{n}\overset{\text{def}}{=}\frac{1}{2}g^{nm}\left( \partial
_{i}g_{mj}+\partial _{j}g_{im}-\partial _{m}g_{ij}\right) \text{.}
\label{cc}
\end{equation}%
Substituting the metric tensor components from (\ref{qq}) into (\ref{cc}),
it turns out that the only non-zero connection coefficients are given by,%
\begin{equation}
\left( \Gamma _{12}^{1}\right) ^{k}=-\frac{1}{\sigma _{k}}\text{, }\left(
\Gamma _{11}^{2}\right) ^{k}=\frac{\alpha _{-}\left( r_{k}\right) }{\alpha
_{+}\left( r_{k}\right) }\frac{1}{\sigma _{k}}\text{, }\left( \Gamma
_{22}^{2}\right) ^{k}=-\frac{1}{\sigma _{k}}\text{,}  \label{1a}
\end{equation}%
where $k=1$,.., $l$. Therefore the set of coupled nonlinear ordinary
differential equations satisfied by the geodesic trajectories becomes,%
\begin{equation}
\frac{d^{2}\tilde{\mu}_{k}}{d\tau ^{2}}-\frac{2}{\tilde{\sigma}}\frac{d%
\tilde{\mu}_{k}}{d\tau }\frac{d\tilde{\sigma}_{k}}{d\tau }=0\text{, }\frac{%
d^{2}\tilde{\sigma}_{k}}{d\tau ^{2}}+\frac{\alpha _{-}\left( r_{k}\right) }{%
\alpha _{+}\left( r_{k}\right) }\frac{1}{\tilde{\sigma}}\left( \frac{d\tilde{%
\mu}_{k}}{d\tau }\right) ^{2}-\frac{1}{\tilde{\sigma}_{k}}\left( \frac{d%
\tilde{\sigma}_{k}}{d\tau }\right) ^{2}=0\text{.}  \label{ss}
\end{equation}%
Notice that in the limit of $r_{k}\rightarrow 0$, $\frac{\alpha _{-}\left(
r_{k}\right) }{\alpha _{+}\left( r_{k}\right) }\rightarrow \frac{1}{2}$ and
the system of equations (\ref{ss}) describing the asymptotic behavior of
maximally probable trajectories on the diagonalized manifold becomes the
standard two-dimensional Gaussian system of nonlinear coupled ordinary
differential equations studied in \cite{cafaroPD}. In order to further
simplify the integration of (\ref{ss}), consider the following (invertible)
change of variables,%
\begin{equation}
\left( \tilde{\mu}_{k}\text{, }\tilde{\sigma}_{k}\right) \longrightarrow
\left( \mu _{k}^{\prime }\left( \tilde{\mu}_{k}\text{, }\tilde{\sigma}%
_{k}\right) =\sqrt{\frac{2\alpha _{-}\left( r_{k}\right) }{\alpha _{+}\left(
r_{k}\right) }}\tilde{\mu}_{k}\text{, }\sigma _{k}^{\prime }\left( \tilde{\mu%
}_{k}\text{, }\tilde{\sigma}_{k}\right) =\tilde{\sigma}_{k}\right) \text{. }
\label{33}
\end{equation}%
Substituting (\ref{33}) into (\ref{ss}), the coupled system of nonlinear
differential equations to be integrated becomes,%
\begin{equation}
\frac{d^{2}\mu _{k}^{\prime }}{d\tau ^{2}}-\frac{2}{\sigma ^{\prime }}\frac{%
d\mu _{k}^{\prime }}{d\tau }\frac{d\sigma _{k}^{\prime }}{d\tau }=0\text{, }%
\frac{d^{2}\sigma _{k}^{\prime }}{d\tau ^{2}}+\frac{1}{2\sigma _{k}^{\prime }%
}\left( \frac{d\mu _{k}^{\prime }}{d\tau }\right) ^{2}-\frac{1}{\sigma
_{k}^{\prime }}\left( \frac{d\sigma _{k}^{\prime }}{d\tau }\right) ^{2}=0%
\text{.}  \label{ddd}
\end{equation}%
Integrating (\ref{ddd}) leads to the following geodesic trajectories,%
\begin{equation}
\mu _{k}^{\prime }\left( \tau \right) =\frac{\Xi _{k}^{2}}{2\lambda _{k}}%
\frac{1}{\exp \left( -2\lambda _{k}\tau \right) +\frac{\Xi _{k}^{2}}{%
8\lambda _{k}^{2}}}-4\lambda _{k}\text{, }\sigma _{k}^{\prime }\left( \tau
\right) =\frac{\Xi _{k}\exp \left( -\lambda _{k}\tau \right) }{\exp \left(
-2\lambda _{k}\tau \right) +\frac{\Xi _{k}^{2}}{8\lambda _{k}^{2}}}\text{,}
\label{34}
\end{equation}%
where $\Xi _{k}$ and $\lambda _{k}$ are \textit{real} and \textit{positive}
constants of integration \cite{cafaroPD}. Using (\ref{re}) and (\ref{33}),
we have%
\begin{equation}
\mu _{k}\left( \mu _{k}^{\prime }\text{, }\sigma _{k}^{\prime }\right) 
\overset{\text{def}}{=}\sqrt{\frac{\alpha _{+}\left( r_{k}\right) }{2\alpha
_{-}\left( r_{k}\right) }}\mu _{k}^{\prime }+\sigma _{k}^{\prime }\text{
and, }\sigma _{k}\left( \mu _{k}^{\prime }\text{, }\sigma _{k}^{\prime
}\right) \overset{\text{def}}{=}\frac{1-\sqrt{\Delta \left( r_{k}\right) }}{%
2r_{k}}\sqrt{\frac{\alpha _{+}\left( r_{k}\right) }{2\alpha _{-}\left(
r_{k}\right) }}\mu _{k}^{\prime }+\frac{1+\sqrt{\Delta \left( r_{k}\right) }%
}{2r_{k}}\sigma _{k}^{\prime }\text{.}
\end{equation}%
Notice that our working hypothesis (\ref{1}) is satisfied since we have,%
\begin{equation}
\lim_{\tau \rightarrow \infty }\frac{\tilde{\mu}_{k}\left( \tau \right) }{%
\tilde{\sigma}_{k}\left( \tau \right) }=\left( \sqrt{\frac{\alpha _{+}\left(
r_{k}\right) }{2\alpha _{-}\left( r_{k}\right) }}\right) \cdot \lim_{\tau
\rightarrow \infty }\frac{\mu _{k}^{\prime }\left( \tau \right) }{\sigma
_{k}^{\prime }\left( \tau \right) }\propto \exp \left( -\lambda _{k}\tau
\right) \overset{\tau \rightarrow \infty }{\longrightarrow }0\text{.}
\end{equation}%
Finally, in terms of the original macrovariables $\left( \mu _{k}\text{, }%
\sigma _{k}\right) $, the geodesic trajectories become, 
\begin{equation}
\begin{array}{c}
\mu _{k}\left( \tau \text{; }r_{k}\right) =\sqrt{\frac{\alpha _{+}\left(
r_{k}\right) }{2\alpha _{-}\left( r_{k}\right) }}\left[ \frac{\Xi _{k}^{2}}{%
2\lambda _{k}}\frac{1}{\exp \left( -2\lambda _{k}\tau \right) +\frac{\Xi
_{k}^{2}}{8\lambda _{k}^{2}}}-4\lambda _{k}\right] +\frac{\Xi _{k}\exp
\left( -\lambda _{k}\tau \right) }{\exp \left( -2\lambda _{k}\tau \right) +%
\frac{\Xi _{k}^{2}}{8\lambda _{k}^{2}}}\text{,} \\ 
\\ 
\sigma _{k}\left( \tau \text{; }r_{k}\right) =\frac{1-\sqrt{\Delta \left(
r_{k}\right) }}{2r_{k}}\sqrt{\frac{\alpha _{+}\left( r_{k}\right) }{2\alpha
_{-}\left( r_{k}\right) }}\left[ \frac{\Xi _{k}^{2}}{2\lambda _{k}}\frac{1}{%
\exp \left( -2\lambda _{k}\tau \right) +\frac{\Xi _{k}^{2}}{8\lambda _{k}^{2}%
}}-4\lambda _{k}\right] +\frac{1+\sqrt{\Delta \left( r_{k}\right) }}{2r_{k}}%
\frac{\Xi _{k}\exp \left( -\lambda _{k}\tau \right) }{\exp \left( -2\lambda
_{k}\tau \right) +\frac{\Xi _{k}^{2}}{8\lambda _{k}^{2}}}\text{.}%
\end{array}
\label{GGE}
\end{equation}%
In our probabilistic macroscopic approach to dynamics, the geodesic
trajectories in (\ref{GGE}) represent the maximum probability paths on the $%
2l$-dimensional embedded Gaussian statistical model.

\section{Information Geometric Complexity}

In our information geometric approach a relevant quantity that can be useful
to study the degree of complexity characterizing information-constrained
dynamical models is the information geometrodynamical entropy $\mathcal{S}_{%
\mathcal{M}_{s}}\left( \tau \right) $ (IGE) \cite{cafaroPD}. In what
follows, we will briefly highlight the key-points leading to the
construction of such quantity.

The elements (or points) $\left\{ P\left( X|\Theta \right) \right\} $ of a $%
2l$-dimensional curved statistical manifold $\mathcal{M}_{s}$ are
parametrized using $2l$-real valued variables $\left( \theta ^{1}\text{,..., 
}\theta ^{2l}\right) $,%
\begin{equation}
\mathcal{M}_{s}\overset{\text{def}}{=}\left\{ P\left( X|\Theta \right)
:\Theta =\left( \theta ^{1}\text{,..., }\theta ^{2l}\right) \in \mathcal{D}%
_{\Theta }^{\left( \text{tot}\right) }\right\} \text{.}
\end{equation}%
The set $\mathcal{D}_{\Theta }^{\left( \text{tot}\right) }$ is the entire
parameter space (at disposal) and it is a subset of $%
\mathbb{R}
^{2l}$,%
\begin{equation}
\mathcal{D}_{\Theta }^{\left( \text{tot}\right) }\overset{\text{def}}{=}%
\dbigotimes\limits_{k=1}^{2l}\mathcal{I}_{\theta ^{k}}=\left( \mathcal{I}%
_{\theta ^{1}}\otimes \mathcal{I}_{\theta ^{2}}\text{...}\otimes \mathcal{I}%
_{\theta ^{2l}}\right) \subseteq 
\mathbb{R}
^{2l}\text{,}
\end{equation}%
where $\mathcal{I}_{\theta ^{k}}$ is a subset of $%
\mathbb{R}
$ and represents the entire range of allowable values for the macrovariable $%
\theta ^{k}$. For instance, considering the statistical manifold of
one-dimensional Gaussian probability distributions parametrized as usual in
terms of $\Theta =\left( \mu \text{, }\sigma \right) $, we obtain%
\begin{equation}
\mathcal{D}_{\Theta }^{\left( \text{tot}\right) }=\mathcal{I}_{\mu }\otimes 
\mathcal{I}_{\sigma }=\left[ \left( -\infty \text{, }+\infty \right) \otimes
\left( 0\text{, }+\infty \right) \right] \subseteq 
\mathbb{R}
^{2}\text{.}
\end{equation}%
In the IGAC, we are interested in a probabilistic description of the
evolution of a given system in terms of its correspondent probability
distribution on $\mathcal{M}_{s}$ which is homeomorphic to $\mathcal{D}%
_{\Theta }^{\left( \text{tot}\right) }$. Assume we are interested in the
evolution from $\tau _{\text{initial}}$ to $\tau _{\text{final}}$. Within
the probabilistic description, this turns out to be equivalent to study the
shortest path (or, in terms of the ME method \cite{caticha2, ariel,
caticha-giffin, adom}, the maximally probable path) leading to $\Theta
\left( \tau _{\text{final}}\right) $ from $\Theta \left( \tau _{\text{initial%
}}\right) $.

Is there a way to quantify the "complexity" of such path?\textbf{.} We have
proposed that the IGE $\mathcal{S}_{\mathcal{M}_{s}}\left( \tau \right) $ is
a good complexity quantifier \cite{carlo-tesi, carlo-CSF}. A suitable
indicator of temporal complexity within the IGAC framework is provided by
the \emph{information geometric entropy} (IGE) $\mathcal{S}_{\mathcal{M}%
_{s}}\left( \tau \right) $ \cite{cafaroPD},%
\begin{equation}
\mathcal{S}_{\mathcal{M}_{s}}\left( \tau \right) \overset{\text{def}}{=}\log 
\widetilde{\emph{vol}}\left[ \mathcal{D}_{\Theta }^{\left( \text{geodesic}%
\right) }\left( \tau \right) \right] \text{.}  \label{info}
\end{equation}%
The average dynamical statistical volume $\widetilde{\emph{vol}}\left[ 
\mathcal{D}_{\Theta }^{\left( \text{geodesic}\right) }\left( \tau \right) %
\right] $ is defined as,%
\begin{equation}
\widetilde{\emph{vol}}\left[ \mathcal{D}_{\Theta }^{\left( \text{geodesic}%
\right) }\left( \tau \right) \right] \overset{\text{def}}{=}\lim_{\tau
\rightarrow \infty }\left( \frac{1}{\tau }\int_{0}^{\tau }d\tau ^{\prime }%
\emph{vol}\left[ \mathcal{D}_{\Theta }^{\left( \text{geodesic}\right)
}\left( \tau ^{\prime }\right) \right] \right) \text{,}
\end{equation}%
where the "tilde" symbol denotes the operation of temporal average. The
volume $\emph{vol}\left[ \mathcal{D}_{\Theta }^{\left( \text{geodesic}%
\right) }\left( \tau ^{\prime }\right) \right] $ is given by,%
\begin{equation}
vol\left[ \mathcal{D}_{\Theta }^{\left( \text{geodesic}\right) }\left( \tau
^{\prime }\right) \right] \overset{\text{def}}{=}\int_{\mathcal{D}_{\Theta
}^{\left( \text{geodesic}\right) }\left( \tau ^{\prime }\right) }\rho
_{\left( \mathcal{M}_{s}\text{, }g\right) }\left( \theta ^{1}\text{,..., }%
\theta ^{n}\right) d^{n}\Theta \text{,}  \label{v}
\end{equation}%
where $\rho _{\left( \mathcal{M}_{s}\text{, }g\right) }\left( \theta ^{1}%
\text{,..., }\theta ^{n}\right) $ is the so-called Fisher density and is
equal to the square root of the determinant $g=\left\vert \det \left( g_{\mu
\nu }\right) \right\vert $ of the metric tensor $g_{\mu \nu }\left( \Theta
\right) $, 
\begin{equation}
\rho _{\left( \mathcal{M}_{s}\text{, }g\right) }\left( \theta ^{1}\text{%
,..., }\theta ^{n}\right) \overset{\text{def}}{=}\sqrt{\left\vert g\left(
\left( \theta ^{1}\text{,..., }\theta ^{n}\right) \right) \right\vert }\text{%
.}
\end{equation}%
The integration space $\mathcal{D}_{\Theta }^{\left( \text{geodesic}\right)
}\left( \tau ^{\prime }\right) $ in (\ref{v}) is defined as follows,%
\begin{equation}
\mathcal{D}_{\Theta }^{\left( \text{geodesic}\right) }\left( \tau ^{\prime
}\right) \overset{\text{def}}{=}\left\{ \Theta \equiv \left( \theta ^{1}%
\text{,..., }\theta ^{n}\right) :\theta ^{k}\left( 0\right) \leq \theta
^{k}\leq \theta ^{k}\left( \tau ^{\prime }\right) \right\} \text{,}
\label{is}
\end{equation}%
where $k=1$,.., $n$ and $\theta ^{k}\equiv \theta ^{k}\left( s\right) $ with 
$0\leq s\leq \tau ^{\prime }$ such that,%
\begin{equation}
\frac{d^{2}\theta ^{k}\left( s\right) }{ds^{2}}+\Gamma _{lm}^{k}\frac{%
d\theta ^{l}}{ds}\frac{d\theta ^{m}}{ds}=0\text{.}
\end{equation}%
The integration space $\mathcal{D}_{\Theta }^{\left( \text{geodesic}\right)
}\left( \tau ^{\prime }\right) $ in (\ref{is}) is a $2l$-dimensional
subspace of the whole (permitted) parameter space $\mathcal{D}_{\Theta
}^{\left( \text{tot}\right) }$. The elements of $\mathcal{D}_{\Theta
}^{\left( \text{geodesic}\right) }\left( \tau ^{\prime }\right) $ are the $%
2l $-dimensional macrovariables $\left\{ \Theta \right\} $ whose components $%
\theta ^{k}$ are bounded by specified limits of integration $\theta
^{k}\left( 0\right) $ and $\theta ^{k}\left( \tau ^{\prime }\right) $ with $%
k=1$,.., $2l$. The limits of integration are obtained via integration of the 
$2l$-dimensional set of coupled nonlinear second order ordinary differential
equations characterizing the geodesic equations. Formally, the IGE $\mathcal{%
S}_{\mathcal{M}_{s}}\left( \tau \right) $ is defined in terms of an averaged
parametric ($\tau $ is the parameter) $2l+1$-fold integral over the
multidimensional geodesic paths connecting $\Theta \left( 0\right) $ to $%
\Theta \left( \tau \right) $. In our information geometric approach, the
information geometric complexity $\widetilde{\emph{vol}}\left[ \mathcal{D}%
_{\Theta }^{\left( \text{geodesic}\right) }\left( \tau \right) \right] $
represents a statistical measure of complexity of the macroscopic path $%
\Theta \overset{\text{def}}{=}\Theta \left( \tau \right) $ on $\mathcal{M}%
_{S}$ connecting the initial and final macrostates $\Theta _{i}$ and $\Theta
_{f}$, respectively. The path $\Theta \left( \tau \right) $ is obtained via
integration of the geodesic equation on $\mathcal{M}_{S}$ generated by the
universal ME updating method. At a discrete level, the path $\Theta \left(
\tau \right) $ can be described in terms of an infinite continuos sequence
of intermediate macroscopic states, $\Theta \left( \tau \right) =\left[
\Theta _{i}\text{,..., }\Theta _{\bar{k}-1}\text{, }\Theta _{\bar{k}}\text{, 
}\Theta _{\bar{k}+1}\text{,..., }\Theta _{f}\right] $ with $\Theta
_{j}=\Theta \left( \tau _{j}\right) $, determined via the logarithmic
relative entropy maximization procedure subjected to well-specified
normalization and information constraints. The nature of such constraints
defines the (correlational) structure of the underlying probability
distribution on the particular curved statistical manifold $\mathcal{M}_{S}$%
. In other words, the correlational structure that may emerge into our
information-geometric statistical models has its origin in the valuable
information about the microscopic degrees of freedom of the actual physical
systems. It emerges in the ME maximization procedure via integration of the
geodesic equations defined on $\mathcal{M}_{S}$ and it is finally quantified
in terms of the intuitive notion of volume growth via the information
geometric complexity $\widetilde{\emph{vol}}\left[ \mathcal{D}_{\Theta
}^{\left( \text{geodesic}\right) }\left( \tau \right) \right] $ or, in
entropic terms by the IGE $\mathcal{S}_{\mathcal{M}_{S}}\left( \tau \right) $%
. The information geometric complexity is then interpreted as the volume of
the statistical macrospace explored in the asymptotic limit by the system in
its complex evolution from $\Theta _{i}$ to $\Theta _{f}$. Otherwise, upon a
suitable normalization procedure that makes the information geometric
complexity an adimensional quantity, it represents the number of accessible
macrostates (with coordinates living in the accessed parameter space $%
\mathcal{D}_{\Theta }^{\left( \text{geodesic}\right) }\left( \tau \right) $)
explored by the system in its evolution from $\Theta _{i}$ to $\Theta _{f}$.

For the model defined in (\ref{SM}), $\mathcal{V}_{\mathcal{M}_{s}}\left(
\tau \right) $ becomes,%
\begin{equation}
\mathcal{V}_{\mathcal{M}_{s}}\left( \tau \right) \equiv \widetilde{\emph{vol}%
}\left[ \mathcal{D}_{\Theta }^{\left( \text{geodesic}\right) }\left( \tau
\right) \right] \overset{\text{def}}{=}\frac{1}{\tau }\dint\limits_{0}^{\tau
}d\tau ^{\prime }\left( \underset{\Theta _{i}\left( 0\right) }{\overset{%
\Theta _{f}\left( \tau ^{\prime }\right) }{\int }}\sqrt{g}d^{2l}\Theta
\right) \text{,}  \label{q2}
\end{equation}%
where $g\overset{\text{def}}{=}\det \left[ g_{ij}\left( \Theta \right) %
\right] _{2l\times 2l}$ is the determinant of the block-diagonal matrix in (%
\ref{here}),%
\begin{equation}
g\left( r_{1}\text{,.., }r_{l}\right) =\dprod\limits_{k=1}^{l}\det \left(
M_{2\times 2}^{\left( k\right) }\right) =\dprod\limits_{k=1}^{l}\left[ \frac{%
2-r_{k}^{2}}{\sigma _{k}^{2}}\right] \text{.}
\end{equation}%
The geodesic paths $\Theta \left( \tau \right) =\left( \mu _{1}\left( \tau 
\text{; }r_{1}\right) \text{, }\sigma _{1}\left( \tau \text{; }r_{1}\right) 
\text{;..; }\mu _{l}\left( \tau \text{; }r_{l}\right) \text{, }\sigma
_{l}\left( \tau \text{; }r_{l}\right) \right) $ are given in (\ref{GGE}).
Substituting them into (\ref{q2}), we get,%
\begin{equation}
\mathcal{V}_{\mathcal{M}_{s}}\left( \tau \text{; }\lambda _{1}\text{,.., }%
\lambda _{k}\right) =\dprod\limits_{k=1}^{l}\left\{ \frac{\sqrt{2-r_{k}^{2}}%
}{\tau }\dint\limits^{\tau }\left[ \frac{\Xi _{k}\exp \left( -\lambda
_{k}\tau ^{\prime }\right) -4\lambda _{k}\sqrt{\frac{\alpha _{+}\left(
r_{k}\right) }{2\alpha _{-}\left( r_{k}\right) }}\exp \left( -2\lambda
_{k}\tau ^{\prime }\right) }{\frac{1+\sqrt{\Delta \left( r_{k}\right) }}{%
2r_{k}}\Xi _{k}\exp \left( -\lambda _{k}\tau ^{\prime }\right) -4\lambda _{k}%
\sqrt{\frac{\alpha _{+}\left( r_{k}\right) }{2\alpha _{-}\left( r_{k}\right) 
}}\frac{1-\sqrt{\Delta \left( r_{k}\right) }}{2r_{k}}\exp \left( -2\lambda
_{k}\tau ^{\prime }\right) }\right] d\tau ^{\prime }\right\} \text{.}
\end{equation}%
For the sake of simplicity, let us introduce the following substitutions,%
\begin{equation}
A_{k}\overset{\text{def}}{=}\Xi _{k}\text{, }B_{k}\overset{\text{def}}{=}%
-4\lambda _{k}\sqrt{\frac{\alpha _{+}\left( r_{k}\right) }{2\alpha
_{-}\left( r_{k}\right) }}\text{, }C_{k}\overset{\text{def}}{=}\frac{1+\sqrt{%
\Delta \left( r_{k}\right) }}{2r_{k}}\Xi _{k}\text{, }D_{k}\overset{\text{def%
}}{=}-4\lambda _{k}\sqrt{\frac{\alpha _{+}\left( r_{k}\right) }{2\alpha
_{-}\left( r_{k}\right) }}\frac{1-\sqrt{\Delta \left( r_{k}\right) }}{2r_{k}}%
\text{.}  \label{casa}
\end{equation}%
Then, the integral defining $\mathcal{V}_{\mathcal{M}_{s}}\left( \tau \text{%
; }\lambda _{1}\text{,..,}\lambda _{k}\right) $ becomes,%
\begin{equation}
\mathcal{V}_{\mathcal{M}_{s}}\left( \tau \text{; }\lambda _{1}\text{,..,}%
\lambda _{k}\right) =\dprod\limits_{k=1}^{l}\left\{ \frac{\sqrt{2-r_{k}^{2}}%
}{\tau }\dint\limits^{\tau }\left[ \frac{A_{k}e^{-\lambda _{k}\tau ^{\prime
}}+B_{k}e^{-2\lambda _{k}\tau ^{\prime }}}{C_{k}e^{-\lambda _{k}\tau
^{\prime }}+D_{k}e^{-2\lambda _{k}\tau ^{\prime }}}d\tau ^{\prime }\right]
d\tau ^{\prime }\right\} \text{.}  \label{29}
\end{equation}%
Upon integration, we get%
\begin{equation}
\int^{\tau }\frac{A_{k}e^{-\lambda _{k}\tau ^{\prime }}+B_{k}e^{-2\lambda
_{k}\tau ^{\prime }}}{C_{k}e^{-\lambda _{k}\tau ^{\prime
}}+D_{k}e^{-2\lambda _{k}\tau ^{\prime }}}d\tau ^{\prime }=\allowbreak \frac{%
1}{\lambda _{k}}\left( \frac{A_{k}}{C_{k}}-\frac{B_{k}}{D_{k}}\right) \ln %
\left[ \frac{D_{k}+C_{k}e^{\lambda _{k}\tau }}{D_{k}e^{\lambda _{k}\tau }}%
\right] +\frac{A_{k}}{C_{k}}\tau \overset{\tau \rightarrow \infty }{\approx }%
\allowbreak \frac{1}{\lambda _{k}}\left( \frac{A_{k}}{C_{k}}-\frac{B_{k}}{%
D_{k}}\right) \ln \frac{C_{k}}{D_{k}}+\frac{A_{k}}{C_{k}}\tau \text{,}
\label{17}
\end{equation}%
and substituting (\ref{17}) into (\ref{29}), we obtain%
\begin{equation}
\mathcal{V}_{\mathcal{M}_{s}}\left( \tau \text{; }\lambda _{1}\text{,..,}%
\lambda _{k}\right) =\dprod\limits_{k=1}^{l}\left\{ \sqrt{2-r_{k}^{2}}\left[ 
\frac{A_{k}}{C_{k}}+\frac{1}{\lambda _{k}}\left( \frac{A_{k}}{C_{k}}-\frac{%
B_{k}}{D_{k}}\right) \ln \frac{C_{k}}{D_{k}}\frac{1}{\tau }\right] \right\} 
\text{.}
\end{equation}%
Introducing again the original parameters in (\ref{casa}), we finally get%
\begin{equation}
\mathcal{V}_{\mathcal{M}_{s}}\left( \tau \text{; }\lambda _{1}\text{,..,}%
\lambda _{k}\right) =\dprod\limits_{k=1}^{l}\left\{ \frac{2r_{k}\sqrt{%
2-r_{k}^{2}}}{1+\sqrt{\Delta \left( r_{k}\right) }}+\left[ \frac{\left(
2r_{k}\sqrt{2-r_{k}^{2}}\right) }{\left( 1+\sqrt{\Delta \left( r_{k}\right) }%
\right) \lambda _{k}}-\frac{2r_{k}\sqrt{2-r_{k}^{2}}}{\left( 1-\sqrt{\Delta
\left( r_{k}\right) }\right) \lambda _{k}}\right] \frac{\ln \Sigma \left(
r_{k}\text{, }\lambda _{k}\text{, }\alpha _{\pm }\right) }{\tau }\right\} 
\text{,}  \label{yes}
\end{equation}%
where the strictly positive function $\Sigma \left( r_{k}\text{, }\lambda
_{k}\text{, }\alpha _{\pm }\right) $ is given by,%
\begin{equation}
\Sigma \left( r_{k}\text{, }\lambda _{k}\text{, }\alpha _{\pm }\right) 
\overset{\text{def}}{=}\frac{\frac{1+\sqrt{\Delta \left( r_{k}\right) }}{%
2r_{k}}\Xi _{k}}{-4\lambda _{k}\sqrt{\frac{\alpha _{+}\left( r_{k}\right) }{%
2\alpha _{-}\left( r_{k}\right) }}\frac{1-\sqrt{\Delta \left( r_{k}\right) }%
}{2r_{k}}}>0\text{, }\forall r_{k}\in \left( 0\text{, }1\right) \text{.}
\end{equation}%
Finally, inserting (\ref{yes}) into (\ref{info}), the IGE $\mathcal{S}_{%
\mathcal{M}_{s}}\left( \tau \right) $ becomes,%
\begin{equation}
\mathcal{S}_{\mathcal{M}_{s}}\left( \tau \text{; }\lambda _{1}\text{,..,}%
\lambda _{k}\right) =\sum_{k=1}^{l}\log \left\{ \frac{2r_{k}\sqrt{2-r_{k}^{2}%
}}{1+\sqrt{\Delta \left( r_{k}\right) }}+\left[ \frac{2r_{k}\sqrt{2-r_{k}^{2}%
}}{\left( 1+\sqrt{\Delta \left( r_{k}\right) }\right) \lambda _{k}}-\frac{%
2r_{k}\sqrt{2-r_{k}^{2}}}{\left( 1-\sqrt{\Delta \left( r_{k}\right) }\right)
\lambda _{k}}\right] \frac{\ln \Sigma \left( r_{k}\text{, }\lambda _{k}\text{%
, }\alpha _{\pm }\right) }{\tau }\right\} \text{.}  \label{ige}
\end{equation}%
With a suitable change of notation, equation (\ref{ige}) can be rewritten in
a more elegant way as follows,%
\begin{equation}
\mathcal{S}_{\mathcal{M}_{s}}\left( \tau \text{; }\lambda _{1}\text{,..,}%
\lambda _{k}\right) \overset{\tau \rightarrow \infty }{\sim }%
\sum_{k=1}^{l}\log \left[ \Lambda _{1}\left( r_{k}\right) +\frac{\Lambda
_{2}\left( r_{k}\text{, }\lambda _{k}\right) }{\tau }\right] \text{,}
\end{equation}%
where,%
\begin{equation}
\Lambda _{1}\left( r_{k}\right) \overset{\text{def}}{=}\frac{2r_{k}\sqrt{%
2-r_{k}^{2}}}{1+\sqrt{1+4r_{k}^{2}}}\text{, }\Lambda _{2}\left( r_{k}\text{, 
}\lambda \right) \overset{\text{def}}{=}\frac{\sqrt{\left(
1+4r_{k}^{2}\right) \left( 2-r_{k}^{2}\right) }}{r_{k}}\frac{\ln \Sigma
\left( r_{k}\text{, }\lambda _{k}\text{, }\alpha _{\pm }\right) }{\lambda
_{k}}\text{, }\alpha _{\pm }\left( r_{k}\right) \overset{\text{def}}{=}\frac{%
3\pm \sqrt{1+4r_{k}^{2}}}{2}\text{. }
\end{equation}%
As stated above\textbf{,} $\Sigma \left( r_{k}\text{, }\lambda _{k}\text{, }%
\alpha _{\pm }\right) $ is a strictly positive function of its arguments. It
appears that the introduction of embedding constraints between the
macrovariables of the larger Gaussian statistical model leads to the
emergence of an asymptotic information geometric compression of the explored
statistical macrostates on the embedded configuration manifold $\mathcal{M}%
_{s}$ in its evolution between the initial and final macrostates. This
result, thought for a special (not general) case, leads to interesting
conclusions. The presence of constraints between macroscopic pieces of
relevant information on the microscopic degrees of freedom of a complex
system allows for an information geometric probabilistic description whose
complexity, measured in terms of\textbf{\ }$\mathcal{S}_{\mathcal{M}_{s}}$%
\textbf{\ (}or\textbf{\ }$\mathcal{V}_{\mathcal{M}_{s}}$\textbf{)}, decays
in a power law way. Asymptotically, the complexity reaches a saturation
value characterized solely by the strength of such macroscopic correlational
coefficients. The relevance of such results becomes even more clear if
compared to what happens for the $4l$-dimensional (uncorrelated and larger)
Gaussian statistical model \cite{cafaroPD}. We will mention this comparison
in our final remarks. For $r_{k}=r_{s}$, $\forall k$, $s=1$,..., $l$, the
information geometric entropy $\mathcal{S}_{\mathcal{M}_{s}}\left( \tau 
\text{; }l\text{, }\lambda \text{, }r\right) $ becomes,%
\begin{equation}
\mathcal{S}_{\mathcal{M}_{s}}\left( \tau \text{; }l\text{, }\lambda \text{, }%
r\right) \overset{\tau \rightarrow \infty }{\sim }\log \left[ \Lambda
_{1}\left( r\right) +\frac{\Lambda _{2}\left( r\text{, }\lambda \right) }{%
\tau }\right] ^{l}\text{.}
\end{equation}%
Therefore, the information geometric complexity presents a power law decay
where the power is related to cardinality $l$ of the microscopic degrees of
freedom characterized by correlated pieces of macroscopic information and it
reaches a saturation value quantified by the set of coefficients $\left\{
r_{k}\right\} $.

\section{Jacobi-Levi-Civita Equation and Jacobi Fields}

The Jacobi-Levi-Civita (JLC) equation of geodesic deviation is a complicated
second-order system of linear ordinary differential equations. It describes
the geodesic spread on curved manifolds of a pair of nearby freely falling
particles travelling on trajectories $\Theta ^{\rho }\left( \tau \right) $
and $\Theta ^{\prime \rho }\left( \tau \right) \overset{\text{def}}{=}\Theta
^{\rho }\left( \tau \right) +\delta \Theta ^{\rho }\left( \tau \right) $.
The JLC equation is given by \cite{felice}\textbf{,}%
\begin{equation}
\frac{\mathcal{D}^{2}J^{k}}{\mathcal{D}\tau ^{2}}+\mathcal{R}_{nml}^{k}\frac{%
\partial \Theta ^{n}}{\partial \tau }J^{m}\frac{\partial \Theta ^{l}}{%
\partial \tau }=0\text{, }  \label{jlc}
\end{equation}%
with $k=1$,.., $2l$ and where the covariant derivatives $\frac{\mathcal{D}%
\Theta ^{\mu }\left( \tau \right) }{\mathcal{D}\tau }$ along the curve $%
\Theta ^{\mu }\left( \tau \right) $ are defined as\textbf{,}%
\begin{equation}
\frac{\mathcal{D}\Theta ^{\mu }\left( \tau \right) }{\mathcal{D}\tau }%
\overset{\text{def}}{=}\frac{d\Theta ^{\mu }\left( \tau \right) }{d\tau }%
+\Gamma _{\nu \rho }^{\mu }\frac{d\Theta ^{\rho }}{d\tau }\Theta ^{\nu }%
\text{.}
\end{equation}%
The Jacobi vector field components $J^{k}$ are given by,%
\begin{equation}
J^{k}\equiv \delta _{\lambda _{k}}\Theta ^{k}\overset{\text{def}}{=}\left( 
\frac{\partial \Theta ^{k}\left( \tau \text{; }\lambda _{k}\right) }{%
\partial \lambda _{k}}\right) _{\tau }\delta \lambda _{k}\text{,}
\label{jacobi}
\end{equation}%
and $\mathcal{R}_{\alpha \beta \gamma \delta }$ is the Riemann curvature
tensor defined as \cite{felice},%
\begin{equation}
\mathcal{R}_{\mu \nu \rho }^{\alpha }\overset{\text{def}}{=}\partial _{\nu
}\Gamma _{\mu \rho }^{\alpha }-\partial _{\rho }\Gamma _{\mu \nu }^{\alpha
}+\Gamma _{\beta \nu }^{\alpha }\Gamma _{\mu \rho }^{\beta }-\Gamma _{\beta
\rho }^{\alpha }\Gamma _{\mu \nu }^{\beta }\text{.}
\end{equation}%
In the $2l$-dimensional case $J=\left\{ J^{k}\right\} _{k=1\text{,.., }2l}$
represents how geodesics, in a $1$-parameter family of geodesics, are
separating. The covariant derivative $\frac{\mathcal{D}^{2}J^{\mu }}{%
\mathcal{D}\tau ^{2}}$ in (\ref{jlc}) is defined as \cite{ohanian}, 
\begin{equation}
\frac{\mathcal{D}^{2}J^{\mu }}{\mathcal{D}\tau ^{2}}=\frac{d^{2}J^{\mu }}{%
d\tau ^{2}}+2\Gamma _{\alpha \beta }^{\mu }\frac{dJ^{\alpha }}{d\tau }\frac{%
d\Theta ^{\beta }}{d\tau }+\Gamma _{\alpha \beta }^{\mu }J^{\alpha }\frac{%
d^{2}\Theta ^{\beta }}{d\tau ^{2}}+\Gamma _{\alpha \beta \text{, }\nu }^{\mu
}\frac{d\Theta ^{\nu }}{d\tau }\frac{d\Theta ^{\beta }}{d\tau }J^{\alpha
}+\Gamma _{\alpha \beta }^{\mu }\Gamma _{\rho \sigma }^{\alpha }\frac{%
d\Theta ^{\sigma }}{d\tau }\frac{d\Theta ^{\beta }}{d\tau }J^{\rho }\text{.}
\label{1d}
\end{equation}%
Equation (\ref{jlc}) forms a system of $2l$ coupled ordinary differential
equations \textit{linear} in the components of the deviation vector field (%
\ref{jacobi}) but\textit{\ nonlinear} in derivatives of the metric tensor $%
g_{ij}\left( \Theta \right) $. It describes the linearized geodesic flow:
the linearization ignores the relative velocity of the geodesics. When the
geodesics are neighboring but their relative velocity is arbitrary, the
corresponding geodesic deviation equation is the so-called generalized
Jacobi equation \cite{chicone}. The nonlinearity is due to the existence of
velocity-dependent terms in the system. Neighboring geodesics accelerate
relative to each other with a rate directly measured by the curvature tensor 
$\mathcal{R}_{\alpha \beta \gamma \delta }$.

Multiplying both sides of (\ref{jlc}) by $g_{ij}\left( \Theta \right) $ and
using the standard symmetry properties of the Riemann curvature tensor, the
geodesic deviation equation becomes,%
\begin{equation}
g_{ji}\frac{\mathcal{D}^{2}J^{i}}{\mathcal{D}\tau ^{2}}+\mathcal{R}_{lmkj}%
\frac{\partial \Theta ^{k}}{\partial \tau }J^{m}\frac{\partial \Theta ^{l}}{%
\partial \tau }=0\text{.}  \label{1c}
\end{equation}%
After some algebra, it follows that the only non-zero Riemann tensor
components are given by,%
\begin{equation}
\left( \mathcal{R}_{1212}\right) ^{k}=-\frac{\alpha _{-}\left( r_{k}\right) 
}{a_{1}^{2}\left( r_{k}\right) }\frac{1}{\sigma _{k}^{2}}\text{, }k=1\text{%
,.., }2l\text{. }  \label{1b}
\end{equation}%
In the model considered, the Jacobi field $\tilde{J}_{\mathcal{M}_{s}}$ has $%
2l$-components $\left\{ \tilde{J}^{\mu }\right\} _{\mu =1\text{,.., }2l}$
that can be grouped into $l$-pairs as follows,%
\begin{equation}
\tilde{J}_{\mathcal{M}_{s}}\leftrightarrow \left( \tilde{J}_{\mathcal{M}%
_{s}}^{1}\text{, }\tilde{J}_{\mathcal{M}_{s}}^{2}\text{,.., }\tilde{J}_{%
\mathcal{M}_{s}}^{2l-1}\text{, }\tilde{J}_{\mathcal{M}_{s}}^{2l}\right)
\leftrightarrow \left( \left( \tilde{J}_{\mathcal{M}_{s}}^{1}\text{, }\tilde{%
J}_{\mathcal{M}_{s}}^{2}\right) _{1\text{-pair}}\text{;..; }\left( \text{ }%
\tilde{J}_{\mathcal{M}_{s}}^{2k-1}\text{, }\tilde{J}_{\mathcal{M}%
_{s}}^{2k}\right) _{k\text{-pair}}\text{;..\thinspace ; }\left( \text{ }%
\tilde{J}_{\mathcal{M}_{s}}^{2l-1}\text{, }\tilde{J}_{\mathcal{M}%
_{s}}^{2l}\right) _{l\text{-pair}}\right) \text{,}  \label{jj}
\end{equation}%
where,%
\begin{equation}
\tilde{J}_{\mathcal{M}_{s}}^{2k-1}\overset{\text{def}}{=}\left( \frac{%
\partial \tilde{\mu}_{k}\left( \tau \text{; }\lambda _{k}\right) }{\partial
\lambda _{k}}\right) _{\tau }\delta \lambda _{k}\text{, }\tilde{J}_{\mathcal{%
M}_{s}}^{2k}\overset{\text{def}}{=}\left( \frac{\partial \tilde{\sigma}%
_{k}\left( \tau \text{; }\lambda _{k}\right) }{\partial \lambda _{k}}\right)
_{\tau }\delta \lambda _{k}\text{,}
\end{equation}%
with $k=1$,.., $l$. Similarly, the $2l$-equations of Jacobi-Levi-Civita can
be grouped into $l$-pairs of differential equations with identical
structure. Substituting (\ref{1a}), (\ref{1b}) and (\ref{1d}) into (\ref{1c}%
), after some tedious algebra, the first pair of JLC equations become,%
\begin{eqnarray}
0 &=&\frac{d^{2}\tilde{J}^{2k-1}}{d\tau ^{2}}+2\left( \Gamma
_{12}^{1}\right) ^{k}\frac{d\tilde{\sigma}_{k}}{d\tau }\frac{d\tilde{J}%
^{2k-1}}{d\tau }+2\left( \Gamma _{12}^{1}\right) ^{k}\frac{d\tilde{\mu}_{k}}{%
d\tau }\frac{d\tilde{J}^{2k}}{d\tau }+  \notag \\
&&  \notag \\
&&+\tilde{J}^{2k-1}\left[ \left( \Gamma _{12}^{1}\right) ^{k}\frac{d^{2}%
\tilde{\sigma}_{k}}{d\tau ^{2}}+\left( \partial _{\tilde{\sigma}_{k}}\left(
\Gamma _{12}^{1}\right) ^{k}+\left( \Gamma _{12}^{1}\right) ^{k}\left(
\Gamma _{12}^{1}\right) ^{k}+\frac{\left( R_{1212}\right) ^{k}}{\left(
g_{11}\right) ^{k}}\right) \left( \frac{d\tilde{\sigma}_{k}}{d\tau }\right)
^{2}\right] +  \notag \\
&&  \notag \\
&&+\tilde{J}^{2k}\left[ \left( \Gamma _{12}^{1}\right) ^{k}\frac{d^{2}\tilde{%
\mu}_{k}}{d\tau ^{2}}+\left( \partial _{\tilde{\sigma}_{k}}\left( \Gamma
_{12}^{1}\right) ^{k}+\left( \Gamma _{12}^{1}\right) ^{k}\left( \Gamma
_{12}^{1}\right) ^{k}-\frac{\left( R_{1212}\right) ^{k}}{\left(
g_{11}\right) ^{k}}\right) \frac{d\tilde{\mu}_{k}}{d\tau }\frac{d\tilde{%
\sigma}_{k}}{d\tau }\right] \text{,}  \label{A}
\end{eqnarray}%
and,%
\begin{eqnarray}
0 &=&\frac{d^{2}\tilde{J}^{2k}}{d\tau ^{2}}+2\left( \Gamma _{11}^{2}\right)
^{k}\frac{d\tilde{\mu}_{k}}{d\tau }\frac{dJ^{2k-1}}{d\tau }+2\left( \Gamma
_{22}^{2}\right) ^{k}\frac{d\tilde{\sigma}_{k}}{d\tau }\frac{dJ^{2k}}{d\tau }%
+  \notag \\
&&  \notag \\
&&+\tilde{J}^{2k-1}\left[ \left( \Gamma _{11}^{2}\right) ^{k}\frac{d^{2}%
\tilde{\mu}_{k}}{d\tau ^{2}}+\left( \partial _{\tilde{\sigma}_{k}}\left(
\Gamma _{11}^{2}\right) ^{k}+\left( \Gamma _{11}^{2}\right) ^{k}\left(
\Gamma _{12}^{1}\right) ^{k}+\left( \Gamma _{22}^{2}\right) ^{k}\left(
\Gamma _{11}^{2}\right) ^{k}-\frac{\left( R_{1212}\right) ^{k}}{\left(
g_{22}\right) ^{k}}\right) \frac{d\tilde{\mu}_{k}}{d\tau }\frac{d\tilde{%
\sigma}_{k}}{d\tau }\right] +  \notag \\
&&  \notag \\
&&+\tilde{J}^{2}\left[ \left( \Gamma _{22}^{2}\right) ^{k}\frac{d^{2}\tilde{%
\sigma}_{k}}{d\tau ^{2}}+\left( \partial _{\tilde{\sigma}_{k}}\left( \Gamma
_{22}^{2}\right) ^{k}+\left( \Gamma _{22}^{2}\right) ^{k}\left( \Gamma
_{22}^{2}\right) ^{k}\right) \left( \frac{d\tilde{\sigma}_{k}}{d\tau }%
\right) ^{2}+\left( \left( \Gamma _{11}^{2}\right) ^{k}\left( \Gamma
_{21}^{1}\right) ^{k}+\frac{\left( R_{1212}\right) ^{k}}{\left(
g_{22}\right) ^{k}}\right) \left( \frac{d\tilde{\mu}_{k}}{d\tau }\right) ^{2}%
\right] \text{.}  \label{B}
\end{eqnarray}%
More explicitly, equations (\ref{A}) and (\ref{B}) may be written as,%
\begin{eqnarray}
0 &=&\frac{d^{2}\tilde{J}^{2k-1}}{d\tau ^{2}}-\frac{2}{\tilde{\sigma}_{k}}%
\frac{d\tilde{\sigma}_{k}}{d\tau }\frac{d\tilde{J}^{2k-1}}{d\tau }-\frac{2}{%
\tilde{\sigma}_{k}}\frac{d\tilde{\mu}_{k}}{d\tau }\frac{d\tilde{J}^{2k}}{%
d\tau }+\tilde{J}^{2k-1}\left[ -\frac{1}{\tilde{\sigma}_{k}}\frac{d_{k}^{2}%
\tilde{\sigma}}{d\tau ^{2}}+\frac{1}{\tilde{\sigma}_{k}^{2}}\left( \frac{d%
\tilde{\sigma}_{k}}{d\tau }\right) ^{2}\right] +  \notag \\
&&  \notag \\
&&+\tilde{J}^{2k}\left[ -\frac{1}{\tilde{\sigma}_{k}}\frac{d^{2}\tilde{\mu}%
_{k}}{d\tau ^{2}}+\frac{3}{\tilde{\sigma}_{k}^{2}}\frac{d\tilde{\mu}_{k}}{%
d\tau }\frac{d\tilde{\sigma}_{k}}{d\tau }\right] \text{,}
\end{eqnarray}%
and,%
\begin{eqnarray}
0 &=&\frac{d^{2}\tilde{J}^{2k}}{d\tau ^{2}}+2\frac{\alpha _{-}\left(
r_{k}\right) }{\alpha _{+}\left( r_{k}\right) }\frac{1}{\tilde{\sigma}_{k}}%
\frac{d\tilde{\mu}_{k}}{d\tau }\frac{d\tilde{J}^{2k-1}}{d\tau }-\frac{2}{%
\tilde{\sigma}_{k}}\frac{d\tilde{\sigma}_{k}}{d\tau }\frac{d\tilde{J}^{2k}}{%
d\tau }+\tilde{J}^{2k-1}\left[ \frac{\alpha _{-}\left( r_{k}\right) }{\alpha
_{+}\left( r_{k}\right) }\frac{1}{\tilde{\sigma}_{k}}\frac{d^{2}\tilde{\mu}%
_{k}}{d\tau ^{2}}-2\frac{\alpha _{-}\left( r_{k}\right) }{\alpha _{+}\left(
r_{k}\right) }\frac{1}{\tilde{\sigma}_{k}^{2}}\frac{d\tilde{\mu}_{k}}{d\tau }%
\frac{d\tilde{\sigma}_{k}}{d\tau }\right] +  \notag \\
&&  \notag \\
&&+\tilde{J}^{2k}\left[ -\frac{1}{\tilde{\sigma}_{k}}\frac{d^{2}\tilde{\sigma%
}_{k}}{d\tau ^{2}}+\frac{2}{\tilde{\sigma}_{k}^{2}}\left( \frac{d\tilde{%
\sigma}_{k}}{d\tau }\right) ^{2}-2\frac{\alpha _{-}\left( r_{k}\right) }{%
\alpha _{+}\left( r_{k}\right) }\frac{1}{\tilde{\sigma}_{k}^{2}}\left( \frac{%
d\tilde{\mu}_{k}}{d\tau }\right) ^{2}\right] \text{.}
\end{eqnarray}%
From (\ref{33}) and (\ref{34}), we notice that the asymptotic expansion of $%
\tilde{\sigma}_{k}$ and the first and second derivative of $\tilde{\mu}_{k}$
and $\tilde{\sigma}_{k}$ are given by,%
\begin{eqnarray}
\tilde{\sigma}_{k}\left( \tau \right) &\approx &\frac{8\lambda _{k}^{2}}{\Xi
_{k}}\exp \left( -\lambda _{k}\tau \right) \text{, }\frac{d\tilde{\mu}_{k}}{%
d\tau }\approx \sqrt{\frac{\alpha _{+}\left( r_{k}\right) }{2\alpha
_{-}\left( r_{k}\right) }}\frac{64\lambda _{k}^{4}}{\Xi _{k}^{2}}\exp \left(
-2\lambda _{k}\tau \right) \text{, }\frac{d\tilde{\sigma}_{k}}{d\tau }%
\approx \frac{8\lambda _{k}^{3}}{\Xi _{k}}\exp \left( -\lambda _{k}\tau
\right) \text{, }  \notag \\
&&  \notag \\
\frac{d^{2}\tilde{\mu}_{k}}{d\tau ^{2}} &\approx &\sqrt{\frac{\alpha
_{+}\left( r_{k}\right) }{2\alpha _{-}\left( r_{k}\right) }}\frac{128\lambda
_{k}^{5}}{\Xi _{k}^{2}}\exp \left( -2\lambda _{k}\tau \right) \text{, }\frac{%
d^{2}\tilde{\sigma}_{k}}{d\tau ^{2}}\approx \frac{8\lambda _{k}^{4}}{\Xi _{k}%
}\exp \left( -\lambda _{k}\tau \right) \text{.}  \label{C}
\end{eqnarray}%
Substituting (\ref{C}) into (\ref{A}) and (\ref{B}), keeping only the
leading terms into the asymptotic expansion, the JLC equations to integrate
become,%
\begin{equation}
\frac{d^{2}\tilde{J}^{2k-1}}{d\tau ^{2}}+2\lambda _{k}\frac{d\tilde{J}^{2k-1}%
}{d\tau }-\sqrt{\frac{\alpha _{+}\left( r_{k}\right) }{2\alpha _{-}\left(
r_{k}\right) }}\frac{16\lambda _{k}^{2}}{\Xi _{k}}\exp \left( -\lambda
_{k}\tau \right) \frac{d\tilde{J}^{2k}}{d\tau }-\sqrt{\frac{\alpha
_{+}\left( r_{k}\right) }{2\alpha _{-}\left( r_{k}\right) }}\frac{8\lambda
_{k}^{3}}{\Xi _{k}}\exp \left( -\lambda _{k}\tau \right) \tilde{J}^{2k}=0%
\text{,}  \label{cc1}
\end{equation}%
and,%
\begin{equation}
\frac{d^{2}\tilde{J}^{2k}}{d\tau ^{2}}+\sqrt{\frac{2\alpha _{-}\left(
r_{k}\right) }{\alpha _{+}\left( r_{k}\right) }}\frac{8\lambda _{k}^{2}}{\Xi
_{k}}\exp \left( -\lambda _{k}\tau \right) \frac{d\tilde{J}^{2k-1}}{d\tau }%
+2\lambda _{k}\frac{d\tilde{J}^{2k}}{d\tau }+\lambda _{k}^{2}\tilde{J}^{2k}=0%
\text{.}  \label{cc2}
\end{equation}%
As a working hypothesis, we assume that \cite{cafaroPD},%
\begin{equation}
\underset{\tau \rightarrow \infty }{\lim }\left[ \exp \left( -\lambda
_{k}\tau \right) \frac{d\tilde{J}^{2k-1}}{d\tau }\right] =0\text{, }\underset%
{\tau \rightarrow \infty }{\lim }\left[ \exp \left( -\lambda _{k}\tau
\right) \frac{d\tilde{J}^{2k}}{d\tau }\right] =0\text{,}\underset{\tau
\rightarrow \infty }{\lim }\left[ \exp \left( -\lambda _{k}\tau \right) 
\tilde{J}^{2k}\right] =0\text{.}  \label{wh}
\end{equation}%
In order to prove that our assumptions in (\ref{wh}) are correct, we will
check \textit{a posteriori }their consistency. The geodesic deviation
equations in (\ref{cc1}) and (\ref{cc2}) finally become,%
\begin{equation}
\frac{d^{2}\tilde{J}^{2k-1}}{d\tau ^{2}}+2\lambda _{k}\frac{d\tilde{J}^{2k-1}%
}{d\tau }=0\text{, }\frac{d^{2}\tilde{J}^{2k}}{d\tau ^{2}}+2\lambda _{k}%
\frac{d\tilde{J}^{2k}}{d\tau }+\lambda _{k}^{2}\tilde{J}^{2k}=0\text{.}
\label{u1}
\end{equation}%
Integration of (\ref{u1}) leads to the following asymptotic expressions for $%
\tilde{J}_{\mathcal{M}_{s}}^{2k-1}\left( \tau \right) $ and $\tilde{J}_{%
\mathcal{M}_{s}}^{2k}\left( \tau \right) $,%
\begin{equation}
\tilde{J}_{\mathcal{M}_{s}}^{2k-1}\left( \tau \right) =C_{0}^{\left(
k\right) }+C_{1}^{\left( k\right) }\exp \left( -2\lambda _{k}\tau \right) 
\text{, }\tilde{J}_{\mathcal{M}_{s}}^{2k}\left( \tau \right) =C_{2}^{\left(
k\right) }\exp \left( -\lambda _{k}\tau \right) +C_{3}^{\left( k\right)
}\tau \exp \left( -2\lambda _{k}\tau \right) \text{,}  \label{usa}
\end{equation}%
where $C_{w}^{\left( k\right) }$ are \emph{real} constant of integration
with $w=1$,.., $3$. Notice that conditions (\ref{wh}) are satisfied and
therefore our assumption are compatible with the solutions obtained.

Consider the Jacobi vector field components $\left\{ \tilde{J}^{\mu
}\right\} _{\mu =1\text{,.., }l}$ defined in (\ref{jj}) and its magnitude $%
\tilde{J}_{\mathcal{M}_{s}}$, \ \ 
\begin{equation}
\tilde{J}_{\mathcal{M}_{s}}^{2}\overset{\text{def}}{=}\left( \tilde{J}_{%
\mathcal{M}_{s}}\right) ^{\mu }\left( \tilde{J}_{\mathcal{M}_{s}}\right)
_{\mu }\text{.}  \label{dd}
\end{equation}%
The magnitude $\tilde{J}_{\mathcal{M}_{s}}$ is called the Jacobi field
intensity. Using (\ref{qq}), Equation (\ref{dd}) becomes,%
\begin{equation}
\tilde{J}_{\mathcal{M}_{s}}^{2}\left( \tau \right) \overset{\text{def}}{=}%
\sum_{k=1}^{l}\left\{ \frac{\alpha _{-}\left( r_{k}\right) }{\left[
a_{1}\left( r_{k}\right) \right] ^{2}}\frac{1}{\tilde{\sigma}_{k}^{2}}\left[ 
\tilde{J}_{\mathcal{M}_{s}}^{2k-1}\left( \tau \right) \right] ^{2}+\frac{%
\alpha _{+}\left( r_{k}\right) }{\left[ a_{1}\left( r_{k}\right) \right] ^{2}%
}\frac{1}{\tilde{\sigma}_{k}^{2}}\left[ \tilde{J}_{\mathcal{M}%
_{s}}^{2k}\left( \tau \right) \right] ^{2}\right\} \text{.}  \label{italy}
\end{equation}%
Substituting (\ref{C}) and (\ref{usa}) into (\ref{italy}), keeping only the
leading terms in the asymptotic expansion of $\ \tilde{J}_{\mathcal{M}%
_{s}}^{2}\left( \tau \right) $, we obtain%
\begin{equation}
\tilde{J}_{\mathcal{M}_{s}}^{2}\left( \tau \right) \overset{\tau \rightarrow
\infty }{\approx }\sum_{k=1}^{l}\left[ \frac{\alpha _{-}\left( r_{k}\right) 
}{\left[ a_{1}\left( r_{k}\right) \right] ^{2}}\left( \frac{C_{0}^{\left(
k\right) }\Xi _{k}}{8\lambda _{k}^{2}}\right) ^{2}\exp \left( 2\lambda
_{k}\tau \right) \right] \text{.}  \label{mmm}
\end{equation}%
Let us rewrite the quantity $\tilde{J}_{\mathcal{M}_{s}}^{2}\left( \tau
\right) $ in terms of "attenuation factors" $\mathcal{A}\left( r_{k}\right) $
and elementary quadratic Jacobi vector field components $j_{\mathcal{M}%
_{s}}^{2}\left( \tau \text{; }\lambda _{k}\right) $ given by,%
\begin{equation}
\widetilde{\mathcal{A}}_{k}\left( r_{k}\right) \overset{\text{def}}{=}\frac{%
\alpha _{-}\left( r_{k}\right) }{\left[ a_{1}\left( r_{k}\right) \right] ^{2}%
}=\frac{2r_{k}\left( 3-\sqrt{1+4r_{k}^{2}}\right) }{\left( 1+\sqrt{%
1+4r_{k}^{2}}\right) ^{2}}\text{,}  \label{m}
\end{equation}%
and,%
\begin{equation}
\tilde{j}_{\mathcal{M}_{s}}^{2}\left( \tau \text{; }\lambda _{k}\right) 
\overset{\text{def}}{=}\left( \frac{C_{0}^{\left( k\right) }\Xi _{k}}{%
8\lambda _{k}^{2}}\right) ^{2}\exp \left( 2\lambda _{k}\tau \right) \text{,}
\label{mm}
\end{equation}%
respectively. Notice that $\widetilde{\mathcal{A}}_{k}\left( r_{k}\right) $
is a bounded function of the correlation coefficient $r_{k}\in \left( 0\text{%
, }1\right) $ and its maximum is reached for $\bar{r}_{k}=\sqrt{2-\sqrt{2}}%
\simeq 0.77$,%
\begin{equation}
\widetilde{\mathcal{A}}_{k\text{max}}\left( r_{k}\right) \overset{\text{def}}%
{=}\underset{r_{k}\in \left( 0\text{, }1\right) }{\max }\widetilde{\mathcal{A%
}_{k}}\left( r_{k}\right) =\widetilde{\mathcal{A}}_{k}\left( \bar{r}%
_{k}\right) =3-2\sqrt{2}\simeq 0.17\text{.}  \label{mmmm}
\end{equation}%
Therefore, substituting (\ref{m}) and (\ref{mm}) into (\ref{mmm}) and
considering the boundedness of the attenuation factors in (\ref{mmmm}), the
square of the Jacobi field intensity $\tilde{J}_{\mathcal{M}_{s}}^{2}\left(
\tau \right) $ may be written as,%
\begin{equation}
\tilde{J}_{\mathcal{M}_{s}}^{2}\left( \tau \text{; }\lambda _{1}\text{,.., }%
\lambda _{l}\right) \overset{\tau \rightarrow \infty }{\approx }%
\sum_{k=1}^{l}\widetilde{\mathcal{A}}_{k}\left( r_{k}\right) \tilde{j}_{%
\mathcal{M}_{s}}^{2}\left( \tau \text{; }\lambda _{k}\right) \text{.}
\label{mika1}
\end{equation}%
Let us now consider the asymptotic behavior of $J_{\mathcal{M}_{s}}$, 
\begin{equation}
J_{\mathcal{M}_{s}}^{2}\overset{\text{def}}{=}\sum_{k=1}^{l}\left\{ \frac{1}{%
\sigma _{k}^{2}}\left[ J_{\mathcal{M}_{s}}^{\left( 2k-1\right) }\right] ^{2}+%
\frac{2r_{k}}{\sigma _{k}^{2}}J_{\mathcal{M}_{s}}^{\left( 2k-1\right) }J_{%
\mathcal{M}_{s}}^{\left( 2k\right) }+\frac{2}{\sigma _{k}^{2}}\left[ J_{%
\mathcal{M}_{s}}^{\left( 2k\right) }\right] ^{2}\right\} \text{.}
\label{fff}
\end{equation}%
From (\ref{re}) and (\ref{jacobi}), it follows that the Jacobi field
components $\left( J_{\mathcal{M}_{s}}^{\left( 2k-1\right) }\text{, }J_{%
\mathcal{M}_{s}}^{\left( 2k\right) }\right) $ with $k=1$,.., $l$ are related
to the Jacobi field components $\left( \tilde{J}_{\mathcal{M}_{s}}^{\left(
2k-1\right) }\text{, }\tilde{J}_{\mathcal{M}_{s}}^{\left( 2k\right) }\right) 
$ in the following way,%
\begin{equation}
J_{\mathcal{M}_{s}}^{\left( 2k-1\right) }\overset{\text{def}}{=}\tilde{J}_{%
\mathcal{M}_{s}}^{\left( 2k-1\right) }+\tilde{J}_{\mathcal{M}_{s}}^{\left(
2k\right) }\text{,}  \label{f}
\end{equation}%
and,%
\begin{equation}
J_{\mathcal{M}_{s}}^{\left( 2k\right) }\overset{\text{def}}{=}\frac{1-\sqrt{%
\Delta \left( r_{k}\right) }}{2r_{k}}\tilde{J}_{\mathcal{M}_{s}}^{\left(
2k-1\right) }+\frac{1+\sqrt{\Delta \left( r_{k}\right) }}{2r_{k}}\tilde{J}_{%
\mathcal{M}_{s}}^{\left( 2k\right) }\text{.}  \label{ff}
\end{equation}%
Considering (\ref{usa}) and substituting (\ref{f}) and (\ref{ff}) into (\ref%
{fff}), the asymptotic behavior of $J_{\mathcal{M}_{s}}^{2}$ is given by,%
\begin{equation}
J_{\mathcal{M}_{s}}^{2}\approx \sum_{k=1}^{l}\left\{ \left[ 1+2r_{k}\frac{1-%
\sqrt{\Delta \left( r_{k}\right) }}{2r_{k}}+2\left( \frac{1-\sqrt{\Delta
\left( r_{k}\right) }}{2r_{k}}\right) ^{2}\right] \frac{1}{\sigma _{k}^{2}}%
\left[ \tilde{J}_{\mathcal{M}_{s}}^{\left( 2k-1\right) }\right] ^{2}\right\} 
\text{.}  \label{qui}
\end{equation}%
Substituting (\ref{GGE}) and (\ref{usa}) into (\ref{qui}) and keeping only
the leading terms into the asymptotic expansion, we obtain,%
\begin{equation}
\left( J_{\mathcal{M}_{s}}^{2}\right) _{\text{embedded}}\left( \tau \text{; }%
\lambda _{1}\text{,.., }\lambda _{l}\right) \approx \sum_{k=1}^{l}\mathcal{A}%
_{k}\left( r_{k}\right) j_{\mathcal{M}_{s}}^{2}\left( \tau \text{; }\lambda
_{k}\right) \text{,}  \label{mika3}
\end{equation}%
where $j_{\mathcal{M}_{s}}^{2}\left( \tau \text{; }\lambda _{k}\right) =%
\tilde{j}_{\mathcal{M}_{s}}^{2}\left( \tau \text{; }\lambda _{k}\right) $ is
defined in (\ref{mm}) and the new attenuation function is given by,%
\begin{equation}
\mathcal{A}_{k}\left( r_{k}\right) \overset{\text{def}}{=}\frac{4r_{k}^{2}%
\left[ 1+2r_{k}a_{0}\left( r_{k}\right) +a_{0}^{2}\left( r_{k}\right) \right]
}{\left[ 1+a_{1}\left( r_{k}\right) \right] ^{2}}\text{,}
\end{equation}%
where $a_{0}\left( r_{k}\right) $ and $a_{1}\left( r_{k}\right) $ are
defined in (\ref{joe}). Notice that $\mathcal{A}_{k}\left( r_{k}\right) $ is
a bounded function of the correlation coefficient $r_{k}\in \left( 0\text{, }%
1\right) $ and its maximum is reached for $\bar{r}_{k}\simeq 0.65$,%
\begin{equation}
\mathcal{A}_{k\text{max}}\left( r_{k}\right) \overset{\text{def}}{=}\underset%
{r_{k}\in \left( 0\text{, }1\right) }{\max }\mathcal{A}_{k}\left(
r_{k}\right) =\mathcal{A}_{k}\left( \bar{r}_{k}\right) \simeq 0.15\text{.}
\end{equation}%
In \cite{cafaroIJTP, cafaroPD}, it was shown that in absence of constraints,
the model here considered leads to an asymptotic behavior of the Jacobi
fields given by,%
\begin{equation}
\left( J_{\mathcal{M}_{s}}^{2}\right) _{\text{larger}}\left( \tau \text{; }%
\lambda _{1}\text{,.., }\lambda _{2l}\right) \overset{\tau \rightarrow
\infty }{\approx }\sum_{k=1}^{2l}j_{\mathcal{M}_{s}}^{2}\left( \tau \text{; }%
\lambda _{k}\right) \text{.}  \label{mika23}
\end{equation}%
Therefore, from equations (\ref{mika23}) and (\ref{mika3}), we obtain,%
\begin{equation}
0\leq \left( \frac{\left[ \left( J_{\mathcal{M}_{s}}^{2}\right) _{\text{%
embedded}}\left( \tau \text{; }\lambda _{1}\text{,.., }\lambda _{l}\right) %
\right] _{_{\bar{k}}}}{\left[ \left( J_{\mathcal{M}_{s}}^{2}\right) _{\text{%
larger}}\left( \tau \text{; }\lambda _{1}\text{,.., }\lambda _{l}\right) %
\right] _{_{\bar{k}}}}\right) ^{\frac{1}{2}}\approx \sqrt{\frac{4r_{\bar{k}%
}^{2}\left[ 1+2r_{\bar{k}}a_{0}\left( r_{\bar{k}}\right) +a_{0}^{2}\left( r_{%
\bar{k}}\right) \right] }{\left[ 1+a_{1}\left( r_{\bar{k}}\right) \right]
^{2}}}\lesssim 0.4<1\text{,}
\end{equation}%
We conclude that the appearance of embedding constraints among the Gaussian
statistical macrovariables on the larger $4l$-dimensional curved manifold
leads to an attenuation of the asymptotic exponential divergence of the
Jacobi field intensity on the embedded $2l$-dimensional manifold. This is a
quantitative indication that the information geometric complexity of a
system decreases in the presence of emerging correlational structures.

\section{Final Remarks}

In this article, we characterized the complexity of geodesic paths on a
curved statistical manifold $\mathcal{M}_{s}$ through the asymptotic
computation of the information geometric complexity $\mathcal{V}_{\mathcal{M}%
_{s}}$ and the Jacobi vector field intensity $J_{\mathcal{M}_{s}}$. We
considered a manifold a $2l$-dimensional Gaussian model\textbf{\ }$\mathcal{M%
}_{s}$ reproduced by an appropriate embedding in a larger\textbf{\ }$4l$%
-dimensional Gaussian manifold and endowed with a Fisher-Rao information
metric $g_{\mu \nu }\left( \Theta \right) $ with non-trivial off diagonal
terms. Such terms in the information metric on the embedded manifold emerged
due to the presence of a correlational structure (embedding constraints)
among the statistical variables on the larger manifold and were
characterized by macroscopic correlational coefficients $r_{k}$.\textbf{\ }%
First, we observed a power law decay of the information geometric complexity
at a rate determined by the coefficients $r_{k}$ and concluded that the
non-trivial off diagonal terms lead to the emergence of an asymptotic
information geometric\textbf{\ }\emph{compression}\textbf{\ }of the explored
macrostates $\Theta $ on $\mathcal{M}_{s}$. Finally,\ we observed that the
presence of such embedding constraints lead to an\textbf{\ }\emph{attenuation%
}\textbf{\ }of the asymptotic exponential divergence of the Jacobi vector
field intensity.

The relevance of such results becomes evident when compared to what happens
in the larger Gaussian statistical model (absence of constraints)\textbf{\ }%
\cite{cafaroPD, cafaroIJTP}. In such case, the information geometric entropy
of the $4l$\textbf{-}dimensional (larger and uncorrelated) Gaussian model
increases linearly in time and its complexity diverges exponentially at a
rate determined by\textbf{\ }$\lambda _{k}$\textbf{, }the Lyapunov exponents
of the statistical trajectories of the system \cite{carlo-tesi} and the
Jacobi field intensity diverges exponentially without any attenuation
factor. It seems that our measure of complexity not only can quantify the
degree of chaoticity of a physical system, but it also adequately captures
the correlational structure (relationship between system's components, \cite%
{F98, H86, G86, JPC89}) in its behavior. In the model studied, the emergence
of structure appears in terms of non-trivial off diagonal elements in the
Fisher-Rao information metric. Such structure leads to the information
geometric compression of $\widetilde{\emph{vol}}\left[ \mathcal{D}_{\Theta
}^{\left( \text{geodesic}\right) }\left( \tau \right) \right] $ and, thus,
to a reduction of the complexity of the path leading to $\Theta _{\text{final%
}}$ from $\Theta _{\text{initial}}$.

Information Geometry and Maximum (relative) Entropy methods hold great
promise for solving computational problems of interest in classical and
quantum physics in terms of their probabilistic description on curved
statistical manifolds. Our theoretical formalism allows us to tackle physics
problems through statistical inference and information geometric techniques,
that is Riemannian geometric techniques applied to probability theory. The
macroscopic behavior of an arbitrary complex system is a consequence of the
underlying statistical structure of the microscopic degrees of freedom of
the system being considered.

As a side remark, we point out two more facts: 1) Probabilistic concepts are
naturally incorporated into the fundamentally statistical quantum theory.
Furthermore, describing and understanding the complexity of quantum motion
is still an open problem since our present knowledge on the relations among
complexity, chaoticity and dynamical stochasticity are not satisfactory at
all \cite{C09}; 2) Riemannian geometric tools are currently being used to
characterize the quantum gate complexity in quantum computing \cite%
{nielsen3, nielsen4, brandt1, brandt2}. In \cite{nielsen3}, the problem of
finding quantum circuits was recasted as a geometric problem. It was shown
that finding optimal quantum circuits is essentially equivalent to finding
the shortest path (geodesic) between two points in a certain curved
geometry. In light of these two considerations and in view of the results
obtained thus far, we are confident the work presented here constitutes a
further important step towards the characterization of the dynamical
complexity of microscopically correlated multidimensional Gaussian
statistical models, and other models of relevance in more realistic physical
systems. We hope to extend this approach in the field of Quantum Information
to better understand the connection between quantum entanglement and quantum
complexity \cite{cafaroMPLB, cafaroPA, C09, nielsen, prosen}.

\begin{acknowledgments}
C. C. thanks Sean Alan Ali, Adom Giffin, Akira Inomata, John Kimball, Kevin
Knuth and Carlos Rodriguez for useful discussions on chaos and complexity.
C. C. is especially indebted to Ariel Caticha for important advises and
illuminating comments. This work was supported by the European Community's
Seventh Framework Program FP7/2007-2013 under grant agreement 213681 (CORNER
Project).
\end{acknowledgments}

\appendix

\section{Derivation of the line element}

We derive Equation (\ref{ntfm}). For the sake of clarity, we consider a
two-dimensional Gaussian probability distribution $p\left( x_{1}\text{, }%
x_{2}|\mu _{1}\text{, }\mu _{2}\text{, }\sigma _{1}\right) $ obtained from
the Gaussian distribution $p\left( x_{1}\text{, }x_{2}|\mu _{1}\text{, }\mu
_{2}\text{, }\sigma _{1}\text{, }\sigma _{2}\right) $ setting $\sigma
_{1}=\sigma _{2}$ and defined as,%
\begin{equation}
p\left( x_{1}\text{, }x_{2}|\mu _{1}\text{, }\mu _{2}\text{, }\sigma
_{1}\right) =\frac{1}{2\pi \sigma _{1}^{2}}\exp \left[ -\frac{1}{2\sigma
_{1}^{2}}\left[ \left( x_{1}-\mu _{1}\right) ^{2}+\left( x_{2}-\mu
_{2}\right) ^{2}\right] \right] \text{.}
\end{equation}%
The Fisher-Rao information metric in the three-dimensional statistical
manifold $\left( \mu _{1}\text{, }\mu _{2}\text{, }\sigma _{1}\right) $ is
given by,%
\begin{equation}
ds^{2}=\frac{1}{\sigma _{1}^{2}}\left( d\mu _{1}^{2}+d\mu _{2}^{2}+4d\sigma
_{1}^{2}\right) \text{.}  \label{im1}
\end{equation}%
Consider the two-dimensional submanifold embedded as a slice in the
three-dimensional space defined by the following embedding constraint,%
\begin{equation}
\mu _{2}=\mu _{2}\left( \mu _{1}\text{, }\sigma _{1}\right) \text{.}
\label{embedding}
\end{equation}%
The Gaussian distributions $p\left( x_{1}\text{, }x_{2}|\mu _{1}\text{, }\mu
_{2}\left( \mu _{1}\text{, }\sigma _{1}\right) \text{, }\sigma _{1}\right)
\equiv \tilde{p}\left( x_{1}\text{, }x_{2}|\mu _{1}\text{, }\sigma
_{1}\right) $ belonging to this submanifold are such that $\left\langle
x_{2}\right\rangle =\mu _{2}$ and $\left\langle x_{1}\right\rangle =\mu _{1}$
are not independent, they are related in a peculiar way. From (\ref%
{embedding}), we obtain%
\begin{equation}
d\mu _{2}=\frac{\partial \mu _{2}}{\partial \mu _{1}}d\mu _{1}+\frac{%
\partial \mu _{2}}{\partial \sigma _{1}}d\sigma _{1}\text{,}
\end{equation}%
that is,%
\begin{equation}
d\mu _{2}^{2}=\left( \frac{\partial \mu _{2}}{\partial \mu _{1}}\right)
^{2}d\mu _{1}^{2}+\left( \frac{\partial \mu _{2}}{\partial \sigma _{1}}%
\right) ^{2}d\sigma _{1}^{2}+2\frac{\partial \mu _{2}}{\partial \mu _{1}}%
\frac{\partial \mu _{2}}{\partial \sigma _{1}}d\mu _{1}d\sigma _{1}\text{.}
\label{rel2}
\end{equation}%
Substituting (\ref{rel2}) in (\ref{im1}), the information metric becomes,%
\begin{equation}
ds^{2}=\frac{1}{\sigma _{1}^{2}}\left[ A_{\mu _{1}\mu _{1}}d\mu
_{1}^{2}+2A_{\mu _{1}\sigma _{1}}d\mu _{1}d\sigma _{1}+2A_{\sigma _{1}\sigma
_{1}}d\sigma _{1}^{2}\right] \text{,}  \label{le2}
\end{equation}%
where the coefficients $A_{\mu _{1}\mu _{1}}$, $A_{\mu _{1}\sigma _{1}}$ and 
$A_{\sigma _{1}\sigma _{1}}$ are given by,%
\begin{equation}
A_{\mu _{1}\mu _{1}}\overset{\text{def}}{=}1+\left( \frac{\partial \mu _{2}}{%
\partial \mu _{1}}\right) ^{2}\text{, }A_{\mu _{1}\sigma _{1}}\overset{\text{%
def}}{=}\frac{\partial \mu _{2}}{\partial \mu _{1}}\frac{\partial \mu _{2}}{%
\partial \sigma _{1}}\text{ and, }A_{\sigma _{1}\sigma _{1}}\overset{\text{%
def}}{=}2+\frac{1}{2}\left( \frac{\partial \mu _{2}}{\partial \sigma _{1}}%
\right) ^{2}\text{.}  \label{relazioni}
\end{equation}%
Re-scaling the variables in such a way that $\tilde{\mu}_{1}\overset{\text{%
def}}{=}A_{\mu _{1}\mu _{1}}^{\frac{1}{2}}\mu _{1}$ and $\tilde{\sigma}_{1}%
\overset{\text{def}}{=}A_{\sigma _{1}\sigma _{1}}^{\frac{1}{2}}\sigma _{1}$
and assuming that the coefficients $A_{\mu _{1}\mu _{1}}$, $A_{\mu
_{1}\sigma _{1}}$ and $A_{\sigma _{1}\sigma _{1}}$ are constants, the line
element in (\ref{le2}) becomes%
\begin{equation}
ds^{2}=A_{\sigma _{1}\sigma _{1}}\frac{1}{\tilde{\sigma}_{1}^{2}}\left[ d%
\tilde{\mu}_{1}^{2}+2r_{1}d\tilde{\mu}_{1}d\tilde{\sigma}_{1}+2d\tilde{\sigma%
}_{1}^{2}\right] \text{,}
\end{equation}%
with,%
\begin{equation}
r_{1}\overset{\text{def}}{=}\frac{A_{\mu _{1}\sigma _{1}}}{A_{\mu _{1}\mu
_{1}}^{\frac{1}{2}}A_{\sigma _{1}\sigma _{1}}^{\frac{1}{2}}}=\frac{\frac{%
\partial \mu _{2}}{\partial \mu _{1}}\frac{\partial \mu _{2}}{\partial
\sigma _{1}}}{\left[ 1+\left( \frac{\partial \mu _{2}}{\partial \mu _{1}}%
\right) ^{2}\right] ^{\frac{1}{2}}\left[ 2+\frac{1}{2}\left( \frac{\partial
\mu _{2}}{\partial \sigma _{1}}\right) ^{2}\right] ^{\frac{1}{2}}}\text{.}
\label{17A}
\end{equation}%
From (\ref{relazioni}), it follows that the embedding defining the
two-dimensional submanifold in the larger three-dimensional manifold of
Gaussians parametrized by $\left( \mu _{1}\text{, }\mu _{2}\text{, }\sigma
_{1}\right) $ is given by,%
\begin{equation}
\mu _{2}\left( \mu _{1}\text{, }\sigma _{1}\right) =a_{1}^{\left( 1\right)
}\mu _{1}+a_{2}^{\left( 1\right) }\sigma _{1}=\frac{a_{1}^{\left( 1\right) }%
}{\sqrt{1+\left[ a_{1}^{\left( 1\right) }\right] ^{2}}}\tilde{\mu}_{1}+\frac{%
a_{2}^{\left( 1\right) }}{\sqrt{2+\frac{1}{2}\left[ a_{2}^{\left( 1\right) }%
\right] ^{2}}}\tilde{\sigma}_{1}\text{,}  \label{27}
\end{equation}%
with,%
\begin{equation}
a_{1}^{\left( 1\right) }\overset{\text{def}}{=}\left( A_{\mu _{1}\mu
_{1}}-1\right) ^{\frac{1}{2}}\text{ and, }a_{2}^{\left( 1\right) }\overset{%
\text{def}}{=}\left( 2A_{\sigma _{1}\sigma _{1}}-4\right) ^{\frac{1}{2}}%
\text{. }
\end{equation}%
Finally, using (\ref{27}) and (\ref{17A}), the explicit expression for $%
r_{1} $ becomes%
\begin{equation}
r_{1}=\frac{a_{1}^{\left( 1\right) }a_{2}^{\left( 1\right) }}{\sqrt{1+\left[
a_{1}^{\left( 1\right) }\right] ^{2}}\sqrt{2+\frac{1}{2}\left[ a_{2}^{\left(
1\right) }\right] ^{2}}}\text{.}  \label{RJ}
\end{equation}%
From (\ref{RJ}), it is transparent that the explicit expression for $r_{1}$
depends on the functional parametric form of the embedding\ constraint in (%
\ref{27}).

\end{document}